\documentclass[12pt,preprint]{aastex}
\usepackage{amssymb,amsmath}

\newcommand\pcc{{\,\rm cm}^{-3}}
\newcommand\K{{\;\rm K}}

\newcommand\yr{{\;\rm yr}}

\newcommand\Msun{{\;\rm\,M_\odot}}

\newcommand\kms{{\;\rm km\; s^{-1}}}

\newcommand\pc{{\;\rm\,pc}}

\newcommand\simgt{\lower.5ex\hbox{$\; \buildrel > \over \sim \;$}}
\newcommand\simlt{\lower.5ex\hbox{$\; \buildrel < \over \sim \;$}}

\begin{document}
\title{Implementation of Sink Particles in the \emph{Athena} Code}
\author{Hao Gong\altaffilmark{1} and Eve C.\ Ostriker\altaffilmark{1,2}}
\email{hgong@astro.umd.edu, eco@astro.princeton.edu}
\altaffiltext{1}{Department of Astronomy, University of Maryland, College Park, MD 20742-2421}
\altaffiltext{2}{Department of Astrophysical Science, Princeton 
University, Princeton NJ 08544}

\begin{abstract}
We describe implementation and tests of sink particle algorithms in the Eulerian grid-based
code \emph{Athena}. Introduction of sink particles enables long-term evolution of systems in
which localized collapse occurs, and it is impractical (or unnecessary) to resolve the 
accretion shocks at the centers of collapsing regions. We discuss similarities and differences 
of our methods compared to other implementations of sink particles. Our criteria for sink 
creation are motivated by the properties of the Larson-Penston collapse solution. We use 
standard particle-mesh methods to compute particle and gas gravity together. Accretion of mass 
and momenta onto sinks is computed using fluxes returned by the Riemann solver. A series of 
tests based on previous analytic and numerical collapse solutions is used to validate our method
and implementation. We demonstrate use of our code for applications with a simulation of planar 
converging supersonic turbulent flow, in which multiple cores form and collapse to create sinks;
these sinks continue to interact and accrete from their surroundings over several Myr.
\end{abstract}


\section{Introduction}
Gravitational collapse is a common feature of many gaseous astrophysical systems, and
specialized methods are required in order to follow collapse in time-dependent hydrodynamic
simulations. These numerical issues are particularly important in studies of star formation.
As gravitational collapse develops, material converges to a central point from all directions 
to create a large density peak. For simulations that follow the formation of 
multiple 
self-gravitating
prestellar cores within large-scale clouds, the true structures that form as a consequence of 
core collapse are 
generally so small that the profile surrounding each collapse center becomes
un-resolvable for grid-based codes, even if adaptive mesh refinement (AMR) is adopted.
For example, stellar radii are $\sim 10^{11} {\rm cm}$, whereas that of a giant molecular 
cloud (GMC) is $\sim 10^{20} {\rm cm}$. With a dynamic range $> 10^9$, it is not possible to
spatially resolve a central post-shock protostar at the same time as capturing the
 large scale flows
that lead to its formation, when multiple collapse centers are simultaneously 
present. When the central density in a collapsing region becomes too large, 
gradients in gas pressure and gravity from the central cell to the neighboring cells
cannot be resolved, such that correct mass and momentum fluxes cannot be computed
by the numerical solvers. A simulation cannot continue under these conditions.

The huge dynamic range involved in gravitational collapse can also lead to difficulties due
to the time step restriction by the Courant condition \citep{rich94}. The time-scale set
by self-gravity varies $\propto \rho^{-1/2}$, so that an increase by a factor of $> 10^6$
in density relative to ambient conditions (as would apply within the centers of collapsed
cores) implies a decrease in the time step by a factor $\sim 10^3$. 

A practical way to deal with the above difficulties, for numerical models that are focused 
on large-scale dynamics, is to establish a minimum spatial resolution and introduce sink particles.
When gravitational collapse occurs, the unresolved high density peaks are eliminated from the 
grid and replaced with sink particles. After a sink is created, the material
can flow smoothly toward the center of collapse, with the profile remaining well resolved near the
sink. Subsequent to the creation of sinks, gas and sink particles
are integrated simultaneously, including mutual gravitational forces. Provided that the flow 
onto sinks is supersonic, introducing them will not affect the dynamics of the upstream flow.
Accretion onto sinks should be implemented in a way that conserves total mass and momentum 
of the system.

\citet{bate95} first introduced sink particle techniques in a smoothed particle hydrodynamics (SPH) 
code. \citet{krum04} and \citet{fede10} implemented sink particles in the grid-based codes ORION and FLASH,
respectively, and extensively discuss tests of their methods. In the past several years, similar 
implementations have been made for a number of other well-established 
codes, such as 
ENZO \citep{wang10}, RAMSES \citep{dubo10,teys11}, and 
 GADGET \citep{japp05}. 
\citet{krum04},
\citet{wang10}, \citet{dubo10}, \citet{teys11} adopt the same methodology, 
including
the criteria for creation of sink particles and the accretion rate onto 
sink particles.
\citet{vazq11} used an early version of the implementation 
in FLASH described by \citet{fede10}.
 In these implementations, \citet{krum04}, \citet{fede10}, \citet{wang10}, 
\citet{pado11}, \citet{vazq11} mainly focus on star formation simulations; \citet{dubo10}, 
\citet{teys11} create sink particles to replace super-massive black holes in cosmological simulations.

In this paper, we present details of our implementation and tests of sink particles in 
the grid-based code \emph{Athena} \citep{gard05,gard08,ston08,ston09}. 
In Section 2, we begin by introducing the Eulerian code and outlining our methods for implementing
sink particles. We physically motivate criteria we adopt for creating sinks, and describe our
methods for treating gravity, accretion, merging, and orbit integration of sink particles. In 
Section 3, we present a series of tests of our methods. These include orbits of two particles,
collapse of self-gravitating spheres with a range of initial conditions (including self-similar 
solutions, and converging supersonic flows). We demonstrate Galilean invariance of our methods.
To illustrate the capabilities of our methods for typical applications, in Section 4 we consider
evolution of a turbulent medium with a large-scale supersonic flow converging to create a dense
slab. We follow the fragmentation of the slab into multiple cores, and the subsequent evolution 
of the system as sink particles are created and grow. Finally, Section 5 summarizes our presentation.

\section{Numerical Methodology}
\subsection{\emph{Athena} Code}
For the simulations presented in this paper we use the three-dimensional (3D) code 
\emph{Athena} (\citealt{gard05,gard08,ston08, ston09}). \emph{Athena} is a grid-based
code that uses higher order Godunov methods to evolve the
time-dependent equations of compressible hydrodynamics and magnetodynamics (MHD), 
allowing for self-gravity, radiative heating and cooling, and other microphysics,
on either a Cartesian or cylindrical \citep{skin10} grid. In this paper, we only refer 
to the hydrodynamics and self-gravity capabilities of the code, with Cartesian 
coordinates. The hydrodynamics equations solved are the mass, momentum, and energy
equations:
\begin{equation}\label{cont_eq}
\frac{\partial \rho}{\partial t} + \mathbf{\nabla}
     \cdot (\rho\,\textbf{\em v}) =0,
\end{equation}
\begin{equation}\label{mom_eq}
\frac{\partial (\rho \textbf{{\em v}})}{\partial t} + 
   \mathbf{\nabla} \cdot (\rho \textbf{\em v}\textbf{\em v} + \textbf{P}^{\ast}) = 
 - \mathbf{\nabla}(\Phi + \Phi_{\rm ext}),
\end{equation}
\begin{equation}\label{eng_eq}
\frac{\partial E}{\partial t} + 
   \mathbf{\nabla} \cdot \left[(E+P)\textbf{\em v}\right] = \rho\textbf{\em v} \cdot
   \mathbf{\nabla}(\Phi + \Phi_{\rm ext}),
\end{equation}
and the Poisson equation,
\begin{equation}\label{poisson_eq}
\mathbf{\nabla}^2 \Phi = 4 \pi G \rho,
\end{equation}
where $\textbf{P}^{\ast}$ is a diagonal tensor with $P^{\ast} \equiv P 
\overset{\leftrightarrow}{\rm I}$, and $P$ is the gas pressure, $E$ is the
total energy density 
\begin{equation}\label{eng_den_eq}
E= \frac{P}{\gamma -1} + \frac{1}{2}\rho v^2,
\end{equation}
$\Phi$ is the gravitational potential of the gas, and $\Phi_{\rm ext}$ is an external
gravitational potential. In this paper, we shall consider isothermal flows, in which the
energy equation (\ref{eng_eq}) is replaced by the relation $P = c_s^2 \rho$. Here
$c_s^2 = k T/\mu$ is the square of the isothermal sound speed, for $T$ the temperature
and $\mu$ the mean mass per-particle.

To solve the Poisson equation under varying boundary conditions, we adopt two different
gravity solvers. For periodic boundary conditions in all directions, we use the fast 
Fourier transformation (FFT) method. For boundary conditions that are periodic in-plane
$(x - y)$ and open in the $z$-direction, we use the FFT method developed by \citet{koya09}.
For open boundaries in all directions, we use a new solver based on the method described in 
\citet{hock81}, which employs FFTs in a zero-padded domain eight times as large as the 
computational box (see Appendix).

{\em Athena} also includes static mesh refinement (SMR), but we do not refer to these
capabilities in the current work. Implementation of our sink particle algorithm with SMR 
will be discussed in a future publication.

\subsection{Creation of Sink Particles}
Different criteria for the creation of sink particles have been discussed in the past 
decade. A density threshold \citep{japp05,pado11,krum04} is the simplest criterion for creation 
of a sink particle. High density regions may form in astronomical systems as a consequence
of different physical processes, most importantly strong supersonic shocks and 
gravitational collapse. If a high density region is not gravitationally bound, it might 
subsequently be destroyed by large scale motions that induce rarefactions. Thus, care
must be taken to select an appropriate density threshold, and to include additional
criteria that must be satisfied before creating sink particles \citep{fede10}.

Using AMR simulations, 
\citet{true97} showed that grid-scale numerical noise could grow to cause artificial 
self-gravitating fragmentation if the local Jeans scale ($L_J \equiv c_s (\pi/G \rho)^{1/2}$)
is not resolved by at least four cells. This criterion gives a density threshold 
\begin{equation}\label{rho_tr}
\rho_{\rm Tr} = \frac{\pi}{16} \frac{c_s^2}{G \Delta x^2},
\end{equation}
for $\Delta x$ the simulation cell size.

The implementation of \citet{krum04} sets the threshold density for sink particle creation
in a cell to $\rho_{\rm Tr}$. 
\citet{bane09} and \citet{vazq11} adopt criteria that first checks whether density
exceeds a threshold density and then checks if the zone in question is a local 
gravitational potential minimum. \citet{bate95} and \citet{fede10} perform a series 
of checks including a density threshold check, a converging flow check, a local gravitational 
potential minimum check, and additional checks that evaluate whether a region is strongly
self-gravitating. 
The converging-flow check of \citet{fede10} requires that 
the flow surrounding a candidate zone is converging 
along all directions, stricter than the condition
$\nabla \cdot \textbf{\em v} < 0$.

For simulations involving gravitational collapse, a sink particle should only be created
at the center of a region that is collapsing. As the sink particle creation criteria
of \citet{krum04} do not specifically limit particle creation to a single collapse center, 
they adopt the approach of merging the spurious sink particles within a given collapsing 
region so that one final sink particle is created inside each potential well.
The extra criteria--- checks for a local potential minimum or a gravitationally bound state 
---adopted by \citet{bane09}, \citet{fede10} and \citet{vazq11} further limit the 
initial creation of sink particles, so that only one sink particle is created for each local 
potential minimum. We shall adopt similar criteria to \citet{fede10} to ensure that single
sink particles are created inside regions that are collapsing. 

\subsubsection{Density threshold}
Our choice of density threshold, $\rho_{\rm thr}$, for sink particle creation is motivated 
by the well-known solution for self-gravitating collapse first obtained by \citet{lars69} 
and \citet{pens69} (hereafter LP). For collapse of an initially-static, 
gravitationally-unstable isothermal sphere, LP showed that a singular density profile
\begin{equation}\label{lp_den}
\rho_{\rm _{LP}}(r)= \frac{8.86 c_s^2}{4\pi G r^2}
\end{equation}
is reached. Numerical studies with a range of initial conditions have shown
that the LP asymptotic solution is in fact an ``attractor'' for isothermal core collapse, no 
matter how the collapse is initiated \citep{bode68,lars69,pens69,hunt77, fost93, ogin99,
henn03,moto03,voro05,gome07,burk09}. Some of the above models start from static unstable 
configurations, and others from static, stable 
configurations that are subjected to an imposed compression, either from enhanced 
external pressure or a converging velocity field, or a core-core collision. These models
all show that collapse starts from outside and propagates in, and that the central
velocity is comparable to the value $-3.28 c_s$ at the time of
singularity formation, when the density profile approaches the inverse-square
LP asymptotic solution $\rho_{\rm LP}$. \citet{gong09,gong11} showed that the collapse
of cores forming inside converging supersonic flows also approaches the
LP solution, whether the flow is spherically symmetric or is turbulent, with no special
symmetry. \citet{gong09,gong11}
also showed that the duration of collapse (starting when the core is $\sim 10$ times
the ambient density) is typically a free-fall time at the mean core density.

When collapse occurs, the LP profile would result in a density
\begin{equation}\label{lp_rho}
\rho_{\rm _{LP}}(0.5 \Delta x) = \frac{8.86}{\pi} \frac{c_s^2}{G \Delta x^2}.
\end{equation}
at a distance $\Delta x/2$ from the center. We take this as the density threshold
($\rho_{\rm thr}$) for sink
particle creation. We note that this value is 14.4 times the value $\rho_{\rm Tr}$ of 
Equation (\ref{rho_tr}). A potential concern is that this might lead to artificial
fragmentation. However, we find (see section 4.3) that using the threshold from 
Equation (\ref{lp_rho}), no additional particles are created compared to cases in which
we instead adopt Equation (\ref{rho_tr}) ($\rho_{\rm Tr}$) for the sink particle
density threshold. After a sink particle is created, all cells 
centered at $r < 2 \Delta x$ become part of a ``sink region'' (see below), such that the 
density of cells exterior to the sink region satisfy the Truelove criterion, with 
$\rho_{\rm LP}(2 \Delta x) = 8.86 c_s^2/(16 \pi G \Delta x^2) < \rho_{\rm Tr}$.

Our standard choice of density threshold is given by Equation (\ref{lp_rho}). However,
we have also tested other choices, as discussed in Section 4.3, and found similar results.

\subsubsection{Control volume}
Surrounding each sink particle is a sink control volume where the gas flow
cannot be resolved. As a sink particle moves from one cell to another, the control 
volume moves with the sink particle, such that the sink particle is always located within
the central zone of the cubic control volume. The sink control volume is generally set to
$(3\Delta x)^3$ although we find similar results for value $(5 \Delta x)^3$ and
$(7\Delta x)^3$. The effective radius of the control volume is $\sim r_{\rm ctrl}$, with
$r_{\rm ctrl} \equiv 1.5 \Delta x$ for a control volume $(3 \Delta x)^3$. As
gravity is unresolved at the same scale, the same control volume is adopted for the
particle-mesh algorithm we use to compute the sink particle's gravity (see section
2.4). Once a control volume has been created,
it acts similar to ghost zones bounding the simulation domain. At every 
time step, 
the density, momentum, and energy of the cells inside each control volume are 
reset using outflow boundary conditions from the active grid (i.e. via 
extrapolation from surrounding non-sink zones).

\subsubsection{Gravitational potential minimum check}
If the density of a cell with integer indexes $(i,j,k)$ exceeds $\rho_{\rm thr}$, 
and its distance to any existing sink particle is larger than $2 r_{\rm ctrl}$,
a ``temporary'' control volume is created surrounding it. This temporary control volume
has same size as the real control volume for a sink particle.
We then check if the central cell is a potential minimum inside this control volume \citep{fede10}.
If the potential minimum test is satisfied, we apply further tests.

\subsubsection{Converging flow check}
As in \citet{fede10} but less restrictive, 
we test whether the candidate sink cell at 
the center of the control volume is surrounded by a converging flow: 
$\nabla \cdot \textbf{\em v} < 0$. Under most circumstances, the tests for
high density, a gravitational potential minimum, and a converging flow 
would guarantee 
the region surrounding this cell is under gravitational collapse. 
However, under the
special circumstance of a strong shock produced by a converging flow, 
all of the above
criteria might be met, but a region would still disperse if the converging 
flow were not
sustained until the region becomes gravitationally bound \citep{fede10}; this 
would occur when the volume of the high-density post-shock region becomes
sufficiently large.

\subsubsection{Gravitationally bound state check}
As a last check before sink particle creation, we test whether the total energy
inside the temporary control volume is negative \citep{fede10}:
\begin{equation}\label{e_total}
E_{\rm grav} + E_{\rm th}+ E_{\rm kin} < 0,
\end{equation}
where $E_{\rm grav}$ is the gravitational potential energy, $E_{\rm th}$ is the thermal
energy, and $E_{\rm kin}$ is the kinetic energy.

The control-volume gravitational energy is calculated as
\begin{equation}\label{e_grav}
E_{\rm grav} = \sum_{ijk} \rho(i,j,k) \Delta \Phi(i,j,k),
\end{equation}
where $\Delta \Phi(i,j,k) = \Phi(i,j,k) - \Phi_0$ is the potential difference between 
the cell with integer index $(i,j,k)$ and the potential at the ``edge'' of the local potential 
well, $\Phi_0$. For $\Phi_0$, we compute the average value of the potential in all zones
immediately outside the temporary control volume. 

The thermal energy and kinetic energy are calculated as follows:
\begin{equation}\label{k_th}
E_{\rm th} = \frac{3}{2}\sum_{ijk} \rho(i,j,k) c_s^2(i,j,k),
\end{equation}

\begin{equation}\label{k_ke}
E_{\rm kin} = \frac{1}{2}\sum_{ijk} \rho(i,j,k)|\textbf{\em v}(i,j,k)-\textbf{\em v}_{\rm cm}|^2;
\end{equation}
here $\textbf{\em v}_{\rm cm}$ is the velocity of the center of mass of the control 
volume.

If a cell passes all the checks above, a sink particle is created and a permanent control 
volume is tagged around it. The initial mass and momentum of the sink particle are set by
the sums over all zones within the control volume. 
Note that the cells within the 
sink control volume are not modified either at the moment of 
sink creation or by subsequent accretion because they are 
effectively ghost zones; 
their values are updated at the same time as other boundary 
conditions (see Section 2.2.2).
Also, as noted above, the sink control volume is redefined whenever the 
sink particle moves to 
a new zone within the computational grid. However, the particle can move within a given 
cell without redefining the sink control volume.

\subsection{Gas Accretion Onto Sinks}
A key aspect of any sink particle implementation is to ensure that the gas accretion rate onto sinks 
is accurate. \citet{krum04} use the Bondi-Hoyle accretion formula to approximate the 
accretion rate, with the sound speed and flow velocity set by host cell values. The
density $\rho_{\infty}$ in the Bondi-Hoyle formula is set based on the mean density in a 
local accretion zone (see their section 2.4). \citet{fede10} take a simpler approach,
removing mass from any cell within the control volume to the sink particles if the density in 
that cell exceeds $\rho_{\rm Tr}$. 

Motivated by the concept of the sink region as ``internal'' ghost zones,
in our algorithm the accretion rates of mass and momentum to each sink particle are 
calculated based on the fluxes returned by the Riemann solver at the interfaces between 
the control volume and the surrounding computational grid. Tests to
confirm that the accretion rate agrees with analytic solutions are discussed in Section 3.
As sink particles cross the border of one cell to enter the next cell, the mass and momentum 
differences between the new and old ghost zones are combined with the fluxes returned by
the Riemann solver to conserve the mass and momentum of the whole simulation domain,
including both gas and particles.

\subsection{Integration and Merging of Sink Particles}
Once they have been created, sink particle position and velocities must be integrated in
time. To do this, we use the leapfrog method \citep[e.g.][]{spri05}, with updates over
time $\Delta t$ given by: 
\begin{equation}\label{dkd}
U(\Delta t) = D\left(\frac{\Delta t}{2}\right)\, K(\Delta t)\, 
        D\left(\frac{\Delta t}{2}\right),
\end{equation}
or 
\begin{equation}\label{kdk}
U(\Delta t) = K\left(\frac{\Delta t}{2}\right)\, D(\Delta t)\, 
        K\left(\frac{\Delta t}{2}\right).
\end{equation}
Here $D(\Delta t)$ and $K(\Delta t)$ are the ``drift'' and ``kick'' operators respectively, 
and $D(\Delta t)$ updates a particle's position without changing its momentum, while 
$K(\Delta t)$ does the opposite; $U(\Delta t)$ is the time evolution operator for an interval 
$\Delta t$. Both drift-kick-drift (DKD, Equation (\ref{dkd})) and kick-drift-kick 
(KDK, Equation (\ref{kdk})) schemes are implemented. For all the simulations presented in 
this work, the KDK scheme is adopted because it is superior to the DKD scheme for variable 
time step \citep{spri05}. The time step of the whole simulation is also restricted by the 
sink particle velocities: sink particles cannot travel further than one grid zone in one time step.

Velocity updates of the sink particles are set based on the gravitational field at the particle's 
(smoothed) location. This gravitational field must include a contribution from all the other particles,
as well as from the gas. We use the triangular-shaped-cloud (TSC) scheme \citep{hock81} to 
calculate the gravity produced by each particle as well as the force each particle feels.
In this method the mass of each sink particle is smoothed to the $n_{\rm _{TSC}}^3$ cells 
surrounding it, where $n_{\rm _{TSC}}$ is the number of cells used in one 
direction. The weights along the $x$ direction are expressed as follows:
\begin{equation}\label{pm_weight}
W_l^x(\Delta h) = \left\{\begin{array}{l}
  \frac{(1-2\frac{\Delta h}{\Delta x})^2}{ 2(n_{\rm _{TSC}}-1)^2} , \quad l=-\frac{n_{\rm _{TSC}}-1}{2}\\
      \\
  \frac{2(n_{\rm _{TSC}}+2l-1-2\frac{\Delta h}{\Delta x})}{(n_{\rm _{TSC}}-1)^2}, \quad l=-\frac{n_{\rm _{TSC}}-1}{2}+1, ...,-1\\
      \\
   \frac{2(n_{\rm _{TSC}}-\frac{3}{2}-2\frac{\Delta h}{\Delta x})}{(n_{\rm _{TSC}}-1)^2} , \quad l = 0 \\
      \\
   \frac{2(n_{\rm _{TSC}}-2l-1+2\frac{\Delta h}{\Delta x})}{(n_{\rm _{TSC}}-1)^2}, \quad l=1,...,\frac{n_{\rm _{TSC}}-1}{2}-1\\
      \\
  \frac{(1+2\frac{\Delta h}{\Delta x})^2}{ 2(n_{\rm _{TSC}}-1)^2}. \quad  l=-\frac{n_{\rm _{TSC}}-1}{2}\\
  \end{array}\right.
\end{equation}
Here $\Delta h$ is the $x$-distance between the sink particle and the center of the cell where
it resides along the $x$ dimension. The index $l$ ranges from $-(n_{\rm _{TSC}}-1)/2$ to 
$(n_{\rm _{TSC}}-1)/2$ and the cell with $l = 0$ is where the sink particle resides. Taking 
$n_{\rm _{TSC}} = 3$, Equation (\ref{pm_weight}) reduces the TSC weights described in 
\citet{hock81}: 
\begin{equation}\label{pm_tsc3}
W_{-1}^x = \frac{1}{8}\left(1-\frac{2\Delta h}{\Delta x}\right)^2; \;\;\;
W_0^x = \frac{3}{4} - \left(\frac{\Delta h}{\Delta x}\right)^2; \;\;\;
W_{+1}^x = \frac{1}{8}\left(1+\frac{2\Delta h}{\Delta x}\right)^2. \;\;\;  
\end{equation}
Analogous weights are defined along the $y$ and $z$-dimensions, giving $W_m^y$ and $W_n^z$.

The weight of the mass in any of the $n_{\rm _{TSC}}^3$ zones of the smoothing volume 
surrounding the particle is a product of the weights from the $x$, $y$, and $z$ dimensions.
After the weights of a sink particle are computed, the particle's mass $m_p$ is applied to 
the grid. Within each particle's smoothing volume, the particle's effective density at zone
$(l, m, n)$ is given by $\frac{W^x_l W^y_m W^z_n m_p}{\Delta x \Delta y \Delta z}$. 
Outside of the sink's control volume, the
density is equal to that of the gas. After the combined gas+particle density is defined, the
solution $\Phi$ of the Poisson equation is obtained via FFTs (using the FFTW package, see 
www.fftw.org). For each zone within the sink control volume, the gravitational field 
$\textbf{\em f}_{\rm lmn}$ is computed taking differences of the potential. The gravitational force on 
each sink particle is computed using the same weights as in Equation (\ref{pm_weight}):
\begin{equation}\label{pm_force}
\textbf{\em f} = \sum_l \sum_m \sum_n W_l^x W_m^y W_m^z \textbf{\em f}_{\rm lmn},
\end{equation}
where $l, m$ and $n$ run from $-(n_{\rm _{TSC}}-1)/2$ to $(n_{\rm _{TSC}}-1)/2$.

With a TSC smoothing volume of $n_{\rm _{TSC}}^3$ zones, the effective radius for smoothing
by the TSC algorithm is $r_{\rm _{TSC}} \equiv n_c \Delta x/2$. For $n_{\rm TSC} = 3, 5$ and $7$, 
$r_{\rm _{TSC}}$ is $1.5 \Delta x, 2.5\Delta x$ and $3.5\Delta x$ respectively. 
With larger $r_{\rm _{TSC}}$, sink particles' masses are smoothed to a larger volume, and 
the gravity is softened more. A smaller value of $r_{\rm _{TSC}}$ gives more accurate gravity 
near a particle. However, the hydrodynamic fluxes are better resolved if the radius $r_{\rm ctrl}$
of the sink control volume (see section 2.2.2) is larger. We have tested different combinations of 
$r_{\rm ctrl}$ and 
$r_{\rm _{TSC}}$ with $r_{\rm ctrl} \ge r_{\rm _{TSC}}$ for the test problems described in 
Section 3.1 and 3.2. We find that different $r_{\rm ctrl}$ give nearly the same accretion rate. 
To maximize resolution, we therefore adopt $n_{\rm _{TSC}} = 3$ and $r_{\rm ctrl} = 
r_{\rm _{TSC}} = 1.5\Delta x$ as our standard choice.

Sink particles in our algorithm represent unresolved star/disk/envelope structures. Because
the gas structure and gravity 
is not resolved within the control volume of each sink particle,  
the detailed
interaction of star/disk/envelope systems that collide or pass near each other cannot directly be followed.
We therefore take a conservative approach and merge two sink particles whenever their control
volumes overlap. A new sink particle is created at the center of mass of the two previous sink 
particles, with total mass and momentum conserved. 
We note that provided the effective mass distribution of each sink 
remains spherically-symmetric and concentrated at a very small scale, it would be safe to 
adopt less-strict merger criteria in methods that compute interparticle
forces directly rather than using particle-mesh gravity; in this case 
close particle-particle interactions require sub-time-stepping
\citep[see e.g.][]{krum04,fede10}.
Because torques on the gas are not well
resolved when the sink region has only $3^3$ zones, we do not presently track the spin angular
momentum of the sink particles.

\section{Tests of the Method}
\subsection{Particle Orbits}
\begin{figure}[ht]
\centerline{
\includegraphics[width=0.9\textwidth]{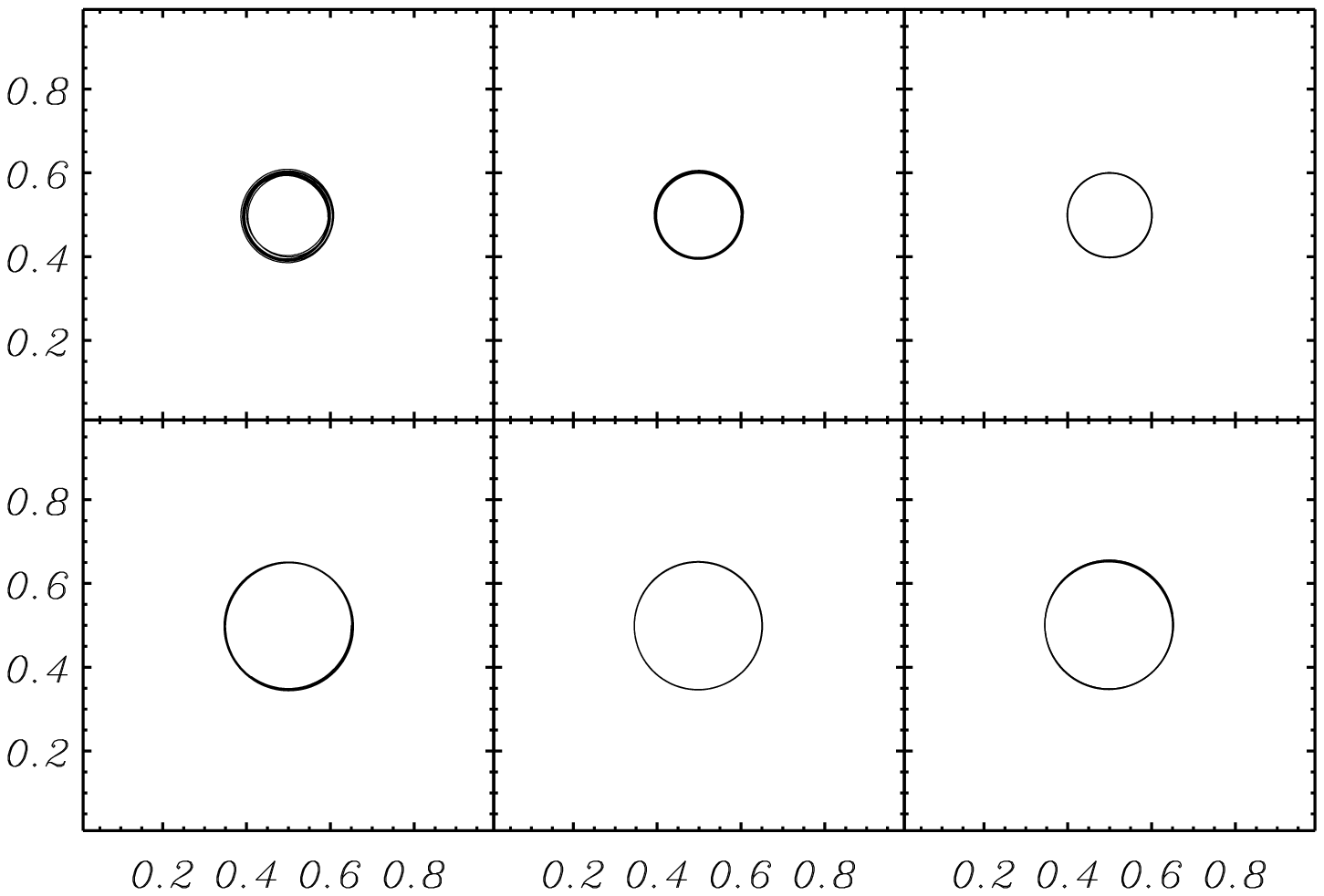}
}
\caption{Circular orbits of two equal mass sink particles orbiting the center of mass 
at different distances, testing our leapfrog particle integrator and TSC particle-mesh Poisson
solver. 
The top panels show orbits of diameter $d/L=0.2$ at box resolutions 
$(L/\Delta x)^3=$ $32^3, 64^3$ and $128^3$
zones from left to right. 
The bottom panels show orbits of diameter $d/L=0.3$ at the same resolutions.
Each panel shows the trajectory of one particle for $10$ orbits.}
\label{fig:circular}
\end{figure}

To test the TSC algorithm and the leapfrog integrator, we consider a problem in which
two particles with equal mass orbit their common center in circular orbits. The initial 
circular
speed of the two particles is $v=\frac{1}{2}\sqrt{2Gm/d}$, where $G$ is the 
gravitational constant, $m$ is the mass of each sink particle, and $d$ is the distance between 
two sink particles.
The gravity between sink particles and gas is disabled, so that sink 
particles only feel gravity from each other. 
Figure \ref{fig:circular} 
shows the trajectory of one particle for $10$ orbits for 
$d/L = 0.2$ (top panels) and 
$d/L=0.3$ (bottom panels) 
with resolutions $32^3, 64^3$ and $128^3$ from left to right. Here, 
$L$ is the simulation box size, corresponding to either 5 or 3.3 times 
the interparticle separation $d$  (which is the 
semimajor axis for the reduced-particle Kepler orbit). 
Vacuum boundary conditions are used for the 
gravity solver.

As long as the gravity is well resolved, the combination of the TSC scheme and
leapfrog integration gives accurate orbits. For $d/L = 0.2$, the 
radius of the orbit is 
$3.2 \Delta x, 6.4 \Delta x$ and $12.8 \Delta x$ for resolution 
$32^3, 64^3$ and $128^3$ 
respectively. For $32^3$ resolution with $r_{\rm ctrl} = 1.5 \Delta x$, the 
cubic smoothing
volumes for each particle are separated by only $2-3$ zones, such that the gravitational
potential is not well approximated by a point mass. For resolution $64^3$ and $128^3$, this 
distance ranges $(8 - 10) \Delta x$ and $(21 - 22) \Delta x$ respectively. We find that
for $64^3$, the orbits are a bit coarse, whereas for $128^3$, the orbits are
nearly perfect. Correspondingly, for the $d/L = 0.3$ cases, 
the orbit radius  is 
$4.8 \Delta x, 9.6 \Delta x$ and 
$19.2 \Delta x$ 
at resolution $32^3, 64^3$ and $128^3$, with circular orbits well represented
at the higher resolutions. We conclude that provided the orbit radius is resolved by $\ge 10$
zones, quite accurate orbits are obtained, with somewhat lower quality orbits at smaller 
separations. However, even for orbit radii as small as 3 zones 
(or $\sim 0.03$pc for typical simulations of core/star formation in molecular 
clouds), orbits are approximately circular. By comparison, we note that \citep{fede11} 
have found that resolution of a vortex in their hydrodynamic simulations on a Cartesian grid 
requires a radius of 15 grid cells.

\subsection{Self-similar Collapse of Isothermal Spheres}
To confirm that our algorithms yield the correct accretion rate, we compare our numerical
results with analytic solutions. In particular, we compare to the family of self-similar
accretions solutions analyzed by \citet{shu77}. The density profile for a generalized 
singular isothermal sphere is 
\begin{equation}\label{rho_shu}
\rho(r) = \frac{Ac_s^2 }{4 \pi G r^2}.
\end{equation}
For $A > 2$, equilibrium is not possible because gravity exceeds the gas 
pressure gradient everywhere; for these initial conditions, a
sphere would globally collapse everywhere. 
The case $A=2$ with zero velocity everywhere 
corresponds to an unstable hydrostatic equilibrium.
\citet{shu77} analyzes a family of self-similar solutions in
which the density profile in the outer parts approaches Equation (\ref{rho_shu}), while 
the inner part approaches a free-fall profile with $\rho \propto r^{-3/2}$. The 
outer part has $v\propto -r^{-1}$, while in the inner part 
the velocity approaches free-fall $v \propto -r^{-1/2}$. For any value of $A$, the central
accretion rate is constant, such that the central mass $M \propto t$, and we can define
a dimensionless accretion rate $\dot{M}/(c_s^3/G)$.

We have tested a series of values of $A$ ranging from $2.0004$ to $4$, the same
as the values in Table 1 of \citet{shu77}. \footnote{
\citet{krum04} show (their Fig 1) a comparison of the numerically-computed density and 
velocity profiles to the semi-analytic collapse solution of  \citet{shu77} for the 
singular isothermal sphere; \citet{fede10} show (their Fig. 10) the density and velocity 
profiles of their numerical solution at several times for the case A=29.3.}
The initial density and 
velocity profiles are the solution of Equations (11)-(12) in \citet{shu77}, 
obtained using 
a four-step Runge-Kutta integration scheme. For code units, we adopt an 
arbitrary density
$\rho_0$, together with length scale, $L_J = c_s (\pi /G \rho_0)^{1/2}$ and 
time scale 
$t_J = L_J/c_s$. 
For all of the tests in this section, the simulation domain size is $(4 L_J)^3$, and the 
resolution is $129^3$. Vacuum boundaries are used for the gravity solver.

The initial radial density and 
velocity profiles are truncated at $r_{\rm max} = 1.5 L_J$. 
Outside this radius, the initial
density is set to $\rho(r = 1.5 L_J)$, 
and the initial velocity is set to zero. 
To convert the self-similar solutions to initial density 
and velocity profiles input to the simulation, 
we choose an initial time $t/t_J = 0.43$ such that in the case $A=2.0004$, the 
initial radius of the expansion wave 
is 11\% of the box size and 29\% of the initial radius of the sphere.
In the case $A=2.0004$, the total mass of the sphere within $r = 1.5 L_J$ is 
$18.9 c_s^3(4\pi G^3 \rho_0)^{-1/2}$.  A sink particle 
is introduced at the center of the sphere at the beginning of the 
simulation, with its 
initial mass set to the mass of the initial profile within $r = r_{\rm ctrl}$ 
(Equations (8), (10) in \citet{shu77}). 
Note that the uniform density outside the sphere leads to 
``outside-in'' collapse because of the imbalance of gravity and 
gas pressure at the truncation radius. 
This process will not affect the initial 
``inside-out'' accretion to the central sink 
particle, however.

\begin{figure}[ht]
\centerline{
\includegraphics[width=0.9\textwidth]{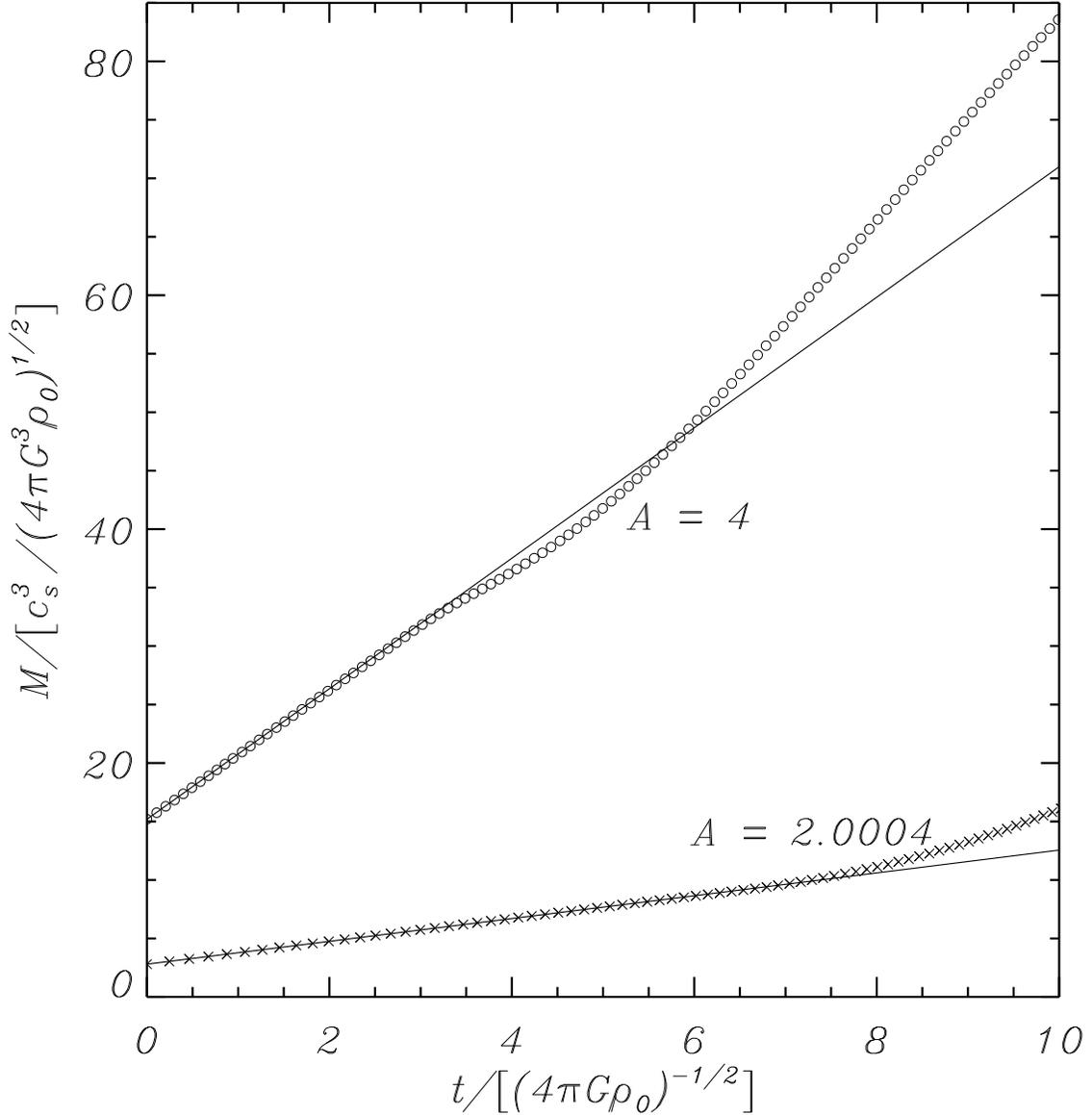}
}
\caption{
Accretion history of the central sink particle for comparison with 
the case $A = 2.0004$ (crosses) and $4$ (open circles) from \citet{shu77}. 
The analytic accretion solutions for self-similar flows 
are shown by the solid lines.  The accretion rates from
our simulations are consistent with the analytical solutions during
early stages.
The enhancement of accretion rates in the later stages is due to the
collapse of the outer part of the initial sphere, 
which is affected by boundary conditions. 
In the units given, the initial sphere masses for the cases $A=2.0004$ and 
$A=4$ are 18.9 and 40.7, respectively.
We fit the linear part of this and other accretion histories to derive the
accretion rates plotted in Figure \ref{fig:singular_accretion}. }
\label{fig:accretion_hist}
\end{figure}

Figure \ref{fig:accretion_hist} shows the accretion history of the
central sink particle for the cases $A = 2.0004$ (crosses) and $4$
(open circles). The solid lines are the analytic solutions from Shu
(1977).  Before effects from the outer boundary conditions begin to
affect the accreting region, the simulation evidently reproduces the
analytical solutions extremely well.  For the case $A=2.0004$, 
the particle mass at $t=6(4 \pi G \rho_0)^{-1/2}$, near 
the end of the linear stage, has reached 
46\% the sphere's initial mass, increasing by a factor three from its 
initial value.
For each value of $A$, we
compute the accretion rate in the simulation using a fit to the sink
particle's mass versus time during the stage when this evolution is
linear.

\begin{figure}[ht]
\centerline{
\includegraphics[width=0.8\textwidth]{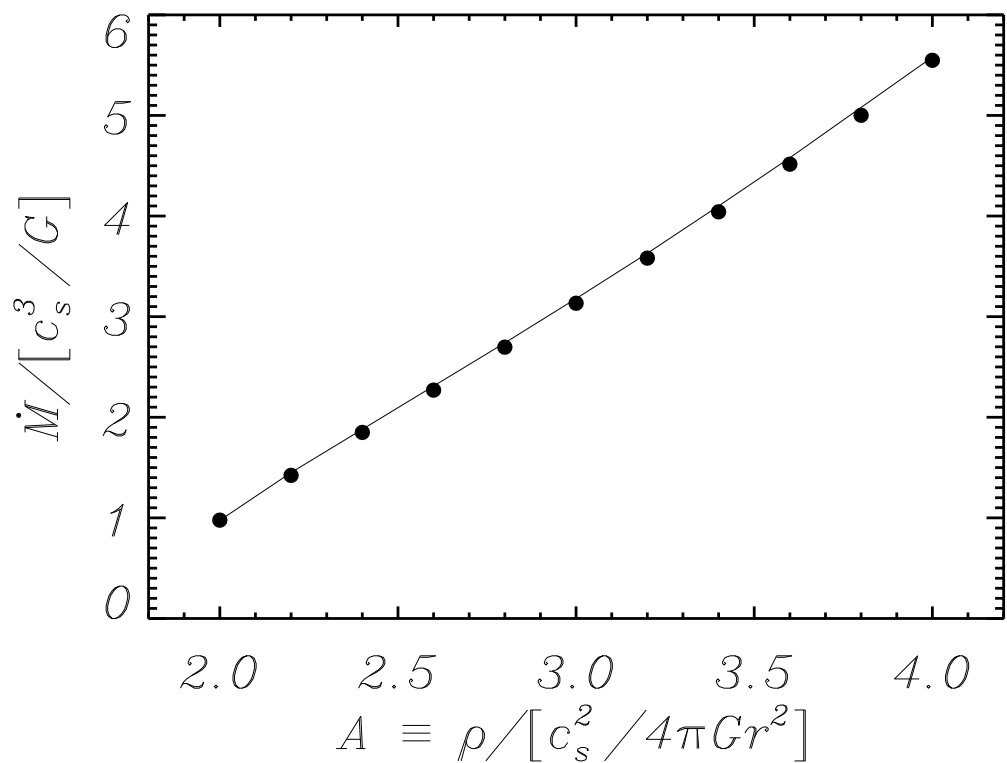}
}
\caption{Accretion rates for self-similar collapse of isothermal spheres with different overdensity
coefficients, $A$ (see Equation (\ref{rho_shu})). The solid line shows the 
analytical accretion rates derived by \citet{shu77}, via direct integration of isothermal fluid 
equations for self-similar models. The solid dots are from accretion rates measured in our
3D simulations.}
\label{fig:singular_accretion}
\end{figure}

Figure \ref{fig:singular_accretion} shows the accretion rates for 
models based on Shu's
solutions with different values of $A$. The line shows the values from the analytic solution 
of \citet{shu77}, and the solid black dots are based on measurements from our numerical 
simulations. The 3D simulations reproduce nearly exactly the predicted accretion rate from the 
analytic solutions.

\begin{figure}[ht]
\centerline{
\includegraphics[width=0.9\textwidth,angle=90]{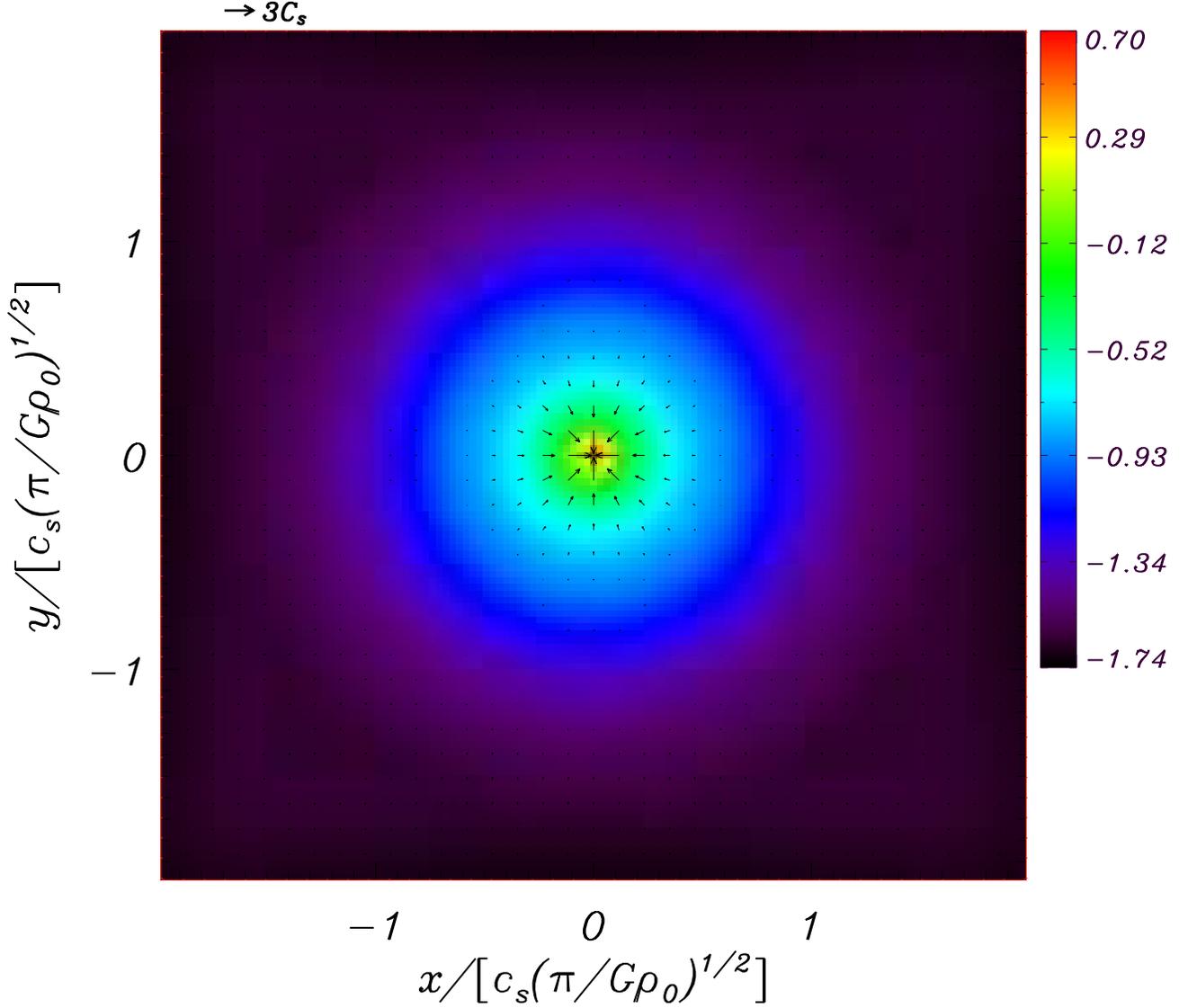}
}
\caption{Density and velocity field cross-section snapshot for the collapsing
near-singular isothermal sphere ($A = 2.0004$). The color scale represents an $x-y$ slice through
the density (${\rm log} \rho$). The direction and length of arrows indicate the 
direction and magnitude of the local velocity, with scale as indicated in the upper left.
The ``expansion wave'' is evident in the plot, with collapse in the inside ($r < 0.67 L_J$)
and a near-static solution in the outside ($r > 0.67 L_J$). The location of the expansion 
wave for the initial conditions was $r = 0.43 L_J$.}
\label{fig:singular_2d_structure}
\end{figure}

\begin{figure}[ht]
\centerline{
\includegraphics[width=0.9\textwidth]{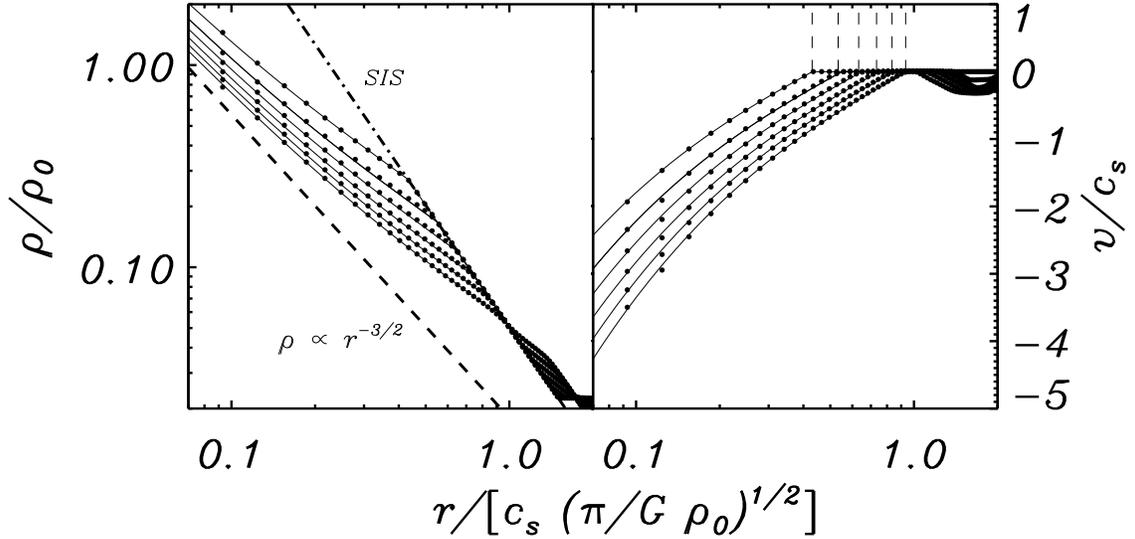}
}
\caption{Radial profiles of density (left panel, filled circles) and velocity (right panel,
 filled circles) for evolution of near-singular ($A = 2.0004$) isothermal sphere. The time interval between profiles 
is constant, equal to $0.1 (\pi/G \rho_0)^{1/2}$, with successive profiles moving to lower 
density and velocity over time.
In both
panels, the solid lines are the 
analytic solutions obtained by integrating 
Equations (11)-(12) in \citet{shu77}. 
For the \citet{shu77} expansion wave 
solution, the outer part of each profile is static (the equilibrium singular isothermal sphere),
and the inner part of each profile approaches free fall, with $\rho \propto r^{-3/2}$ and $v 
\propto r^{-1/2}$. The dot dash line in the left panel shows the singular isothermal 
sphere density profile. The velocity profiles are plotted in linear-log scale to show the 
propagation of the ``expansion wave''. The dashed lines in the right panel indicate the position 
of the ``expansion wave'' front. 
The consistency of our numerical solution with the \citet{shu77} solution 
is evident.
}
\label{fig:singular_rhov_profile}
\end{figure}

Figure \ref{fig:singular_2d_structure} shows a sample cross-section of the density and
velocity field near the center of the collapsing singular sphere for $A = 2.0004$.
Since $A$ is close to $2$,  the simulated sphere recovers Shu's famous ``expansion-wave''
solution: the outer part retains a static singular isothermal equilibrium profile before 
the expansion wave arrives, and the inner part is free-falling towards the center. 
Figure \ref{fig:singular_rhov_profile} shows the radial density profiles and velocity 
profiles from the simulation at different instants during the collapse. The filled circles
are the profiles from the numerical simulation and the solid lines are the corresponding
analytic solutions. Our simulation reproduces the analytic solutions.
The ``expansion wave'' is clearly seen from both the density and velocity profiles.
In both density and velocity profiles, it is seen that gas very near the boundaries collapses, 
and the outer density is slightly altered from $\rho \propto r^{-2}$. This behavior is due 
to conditions near the boundaries, but this process does not affect the collapse in the 
interior.

\subsection{Galilean Invariance of Accretion}
To confirm that the accretion is properly computed for moving particles, we consider tests
in which both the particle and surrounding gas are initialized with a bulk flow
across the grid. The initial conditions of these models are exactly same as the $A = 2.0004$ 
case in Section 3.2, but with an additional uniform bulk flow of speed 
$v_{\rm ad} =  0, 0.5c_s, 1.5c_s$ and $2.5c_s$. 
If the update of the mass and momentum of the sink particle is correct, it 
will continue to move with the surrounding gas sphere at the same velocity. The mass accretion 
rate onto the sink particle should also be the same for different bulk flow speeds.
For this problem, periodic boundaries are adopted for both gas and the gravity solver.
For all of the tests in this section, the simulation size is $(4 L_J)^3$, 
and the resolution is $129^3$.

We note that the boundary conditions in Section 3.2 are open,
different from what we adopt in this section.  Here, we adopt periodic
boundary conditions so that the sphere may pass through the boundary
of the simulation domain (for large Mach number cases).  Because of
the difference in boundary conditions, the accretion rate is
different from that in Section 3.2; however we shall show that 
the accretion rates are all consistent with each other 
for different Mach numbers.

\begin{figure}[ht]
\centerline{
\includegraphics[width=0.9\textwidth]{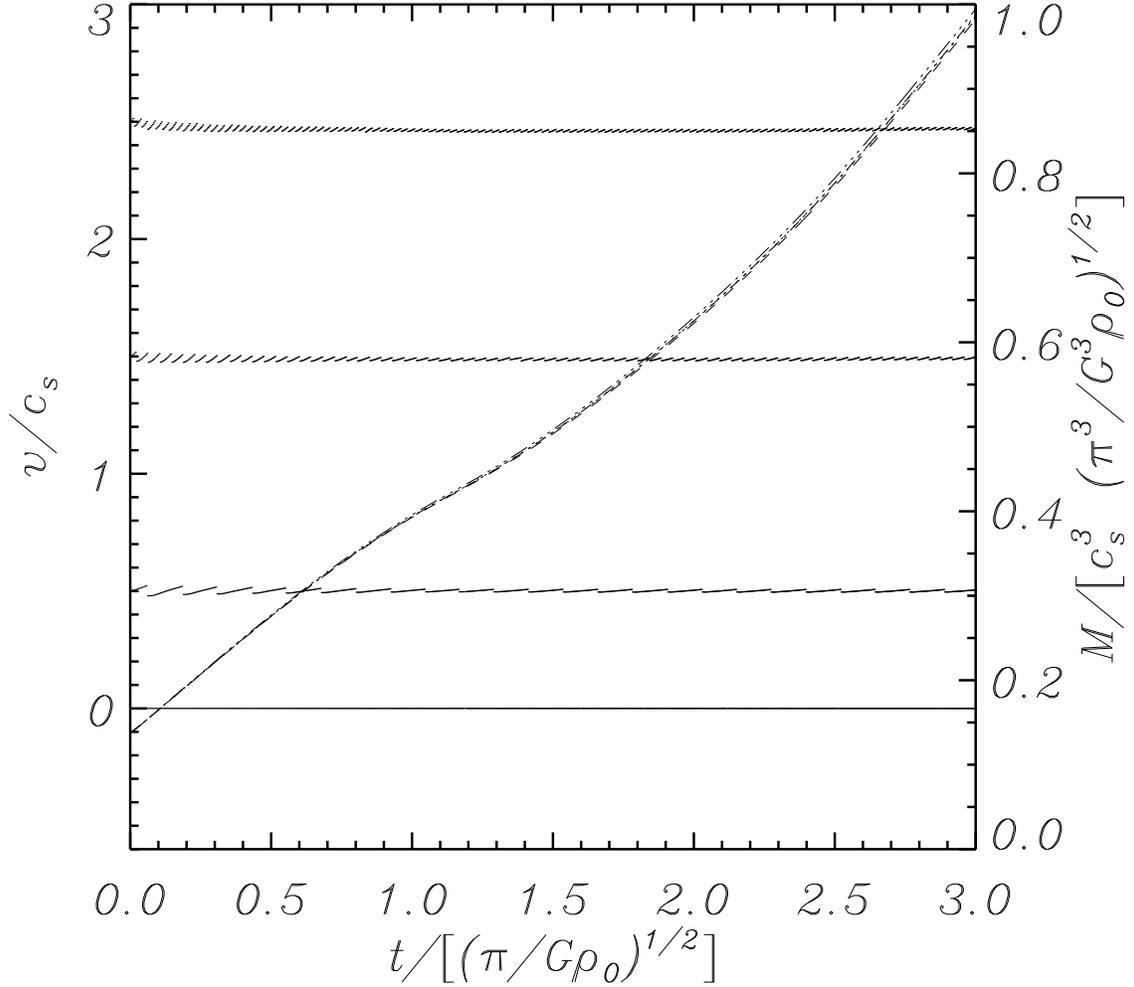}
}
\caption{Tests of accretion onto a sink particle for the collapsing near-singular isothermal 
sphere ($A = 2.0004$) with bulk motion across the grid, demonstrating Galilean invariance. 
Horizontal curves, left scale: the temporal evolution of the speed
of the sink particle for different cases. From bottom to top, the bulk advection speed is
$0, 0.5 c_s$, $1.5 c_s$ and $2.5 c_s$. The time-averaged particle speeds remain constant, 
and equal
to that of the bulk flow. Diagonal curves, right scale: the temporal evolution of the mass 
of the central sink particle for the four different cases. The accretion rate is same for 
static, subsonic, and supersonic advection cases, confirming Galilean invariance of our
algorithms.}
\label{fig:lp_mv}
\end{figure}

Figure \ref{fig:lp_mv} shows the speed and mass of the sink particle versus time, for all tests. 
For the case of zero bulk advection, the speed of
the particle remains zero all the time. For $v_{\rm ad} \neq 0$, the mean speed of the sink 
particle agrees with the bulk speed, with small oscillations. During an interval 
$\Delta x/v_{\rm ad}$, the
time for the sink particle to cross a zone, the control volume does not move. However, as
discussed in Section 2.2.2, the sink control volume must be reset when the particle crosses
a zone boundary. During the time interval $\Delta x/v_{\rm ad}$, the mass and 
momentum fluxes into the control volume are not exactly symmetric, because the gravitational 
potential is not symmetric if the particle is offset from its zone center. The downwind 
material enters the control volume with relatively larger momentum than the upwind material, and 
addition of net positive momentum accelerates the sink particle. The speed of the sink 
particle therefore temporarily increases. As the particle crosses the border of one cell 
to enter the next cell, the sink control volume is shifted, and mass and momentum 
differences from cells entering and leaving the control volume are applied to the 
particle to conserve mass and momentum. As a consequence, the particle's speed is reduced
immediately after crossing a cell edge. At late times, the fractional change in the momentum
becomes small, so the oscillation amplitude drops.

Figure \ref{fig:lp_mv} shows that the mass of the sink particle increases at the same rate 
for all the tests, at varying $v_{\rm ad}$. At early stages, before 
material originating near the sphere's
boundaries accretes to the sink particle, 
differences in the accretion rate are completely 
negligible, and and differences 
remain very small even after material originating near the 
boundaries reaches the center.

We conclude that our sink particle algorithm satisfies Galilean invariance, with the accretion
rate the same whether or not the flow and the grid are in relative motion.

\subsection{Collapse of a Bonnor-Ebert Sphere}
To test our algorithm for the the creation of  a sink particle, together with the accretion
after formation, we run 3D simulations of the evolution of a Bonnor-Ebert sphere 
\citep[hereafter BE]{bonn56,eber55} and compare the results to a 1D spherically-symmetric 
simulation conducted with another code. We consider a BE sphere with a radius slightly 
larger than the critical radius, $r = 1.1 r_{\rm BE,crit}$ for 
$r_{\rm BE,crit} = 0.274 c_s/(\pi / G \rho_e)^{1/2}$, where $\rho_e$ is the 
edge density. The density inside the sphere is everywhere twice as large as the equilibrium 
value. Exterior the sphere, the density is set to $0.001 \rho_e$. The initial velocity
is everywhere set to zero for the initial conditions. The gas boundary conditions are 
outflow. The resolution is $129^3$, and the box size is $[0.301 c_s/(\pi / G \rho_e)^{1/2}]^3$.

\begin{figure}[ht]
\centerline{
\includegraphics[width=0.9\textwidth,angle=90]{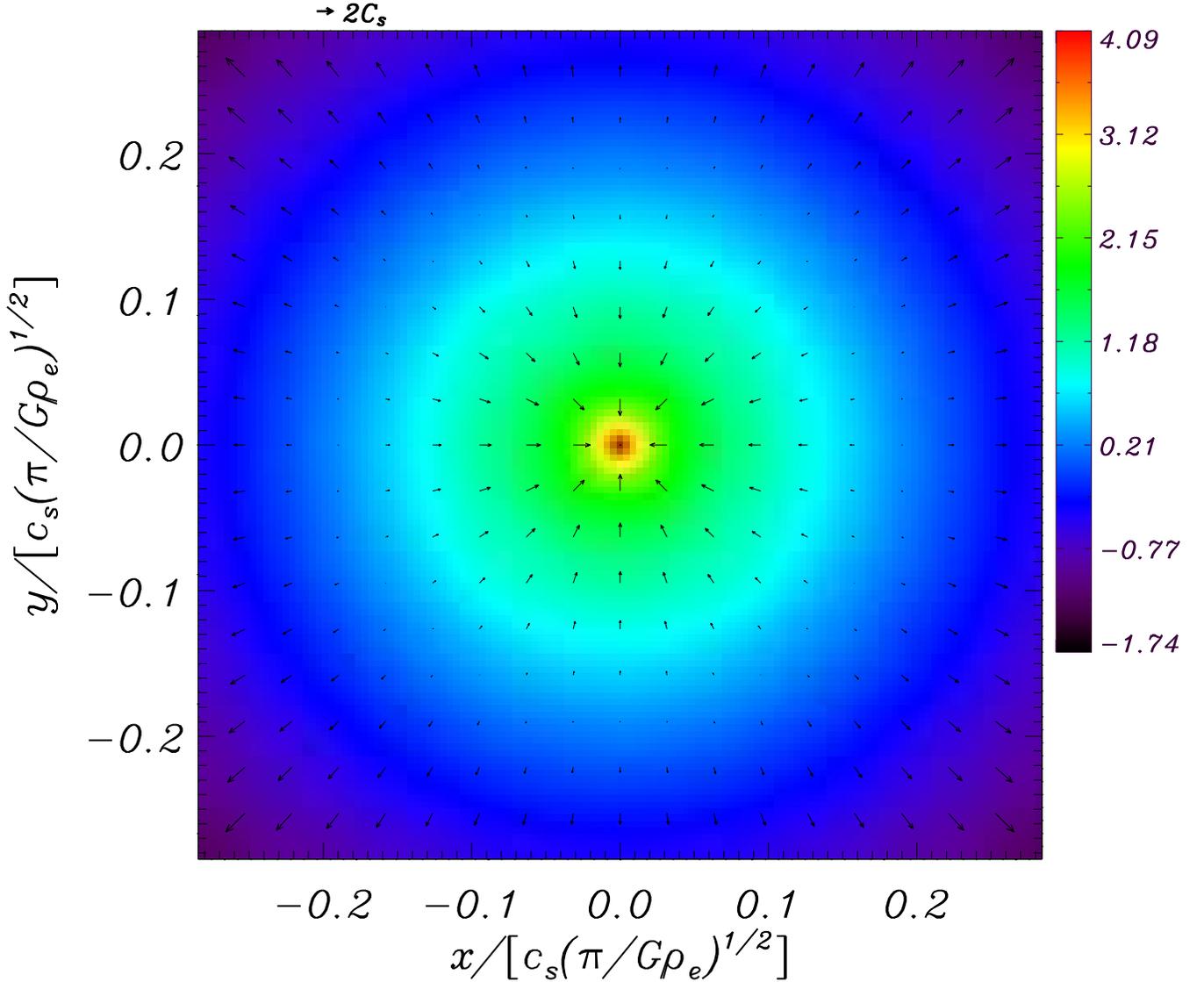}
}
\caption{Density and velocity field cross-section at the instant of singularity for collapse
of a Bonnor-Ebert sphere. The color scale represents an $x-y$ slice through
the volume density (${\rm log} \rho/\rho_e$, for $\rho_e$ the initial density at the outer 
edge of the sphere). The direction and length of arrows indicate the 
direction and magnitude of the local velocity, with scale as indicated in the upper left.
The inner part of the sphere is collapsing, with velocity increasing inward to approach
$-3.4 c_s$. Because we do not have a hot, high-pressure confining medium, the outer
part of the sphere also expands.}
\label{fig:be_2d_structure}
\end{figure}

Figure \ref{fig:be_2d_structure} shows the cross-section of the density and velocity field
across the center of the BE sphere immediately before the central sink particle is created. 
Evidently, the inner part collapses inwards, following the ``outside-in'' collapse described 
by \citet{lars69}. Because the outer boundary of the sphere is not in equilibrium
(for this isothermal simulation, we do not have a hot confining medium), 
the outer part of the sphere expands outwards at the same time.
Only the very late accretion to the sink particle is affected by the early expansion
of the outer parts of the sphere.

\begin{figure}[ht]
\centerline{
\includegraphics[width=0.9\textwidth]{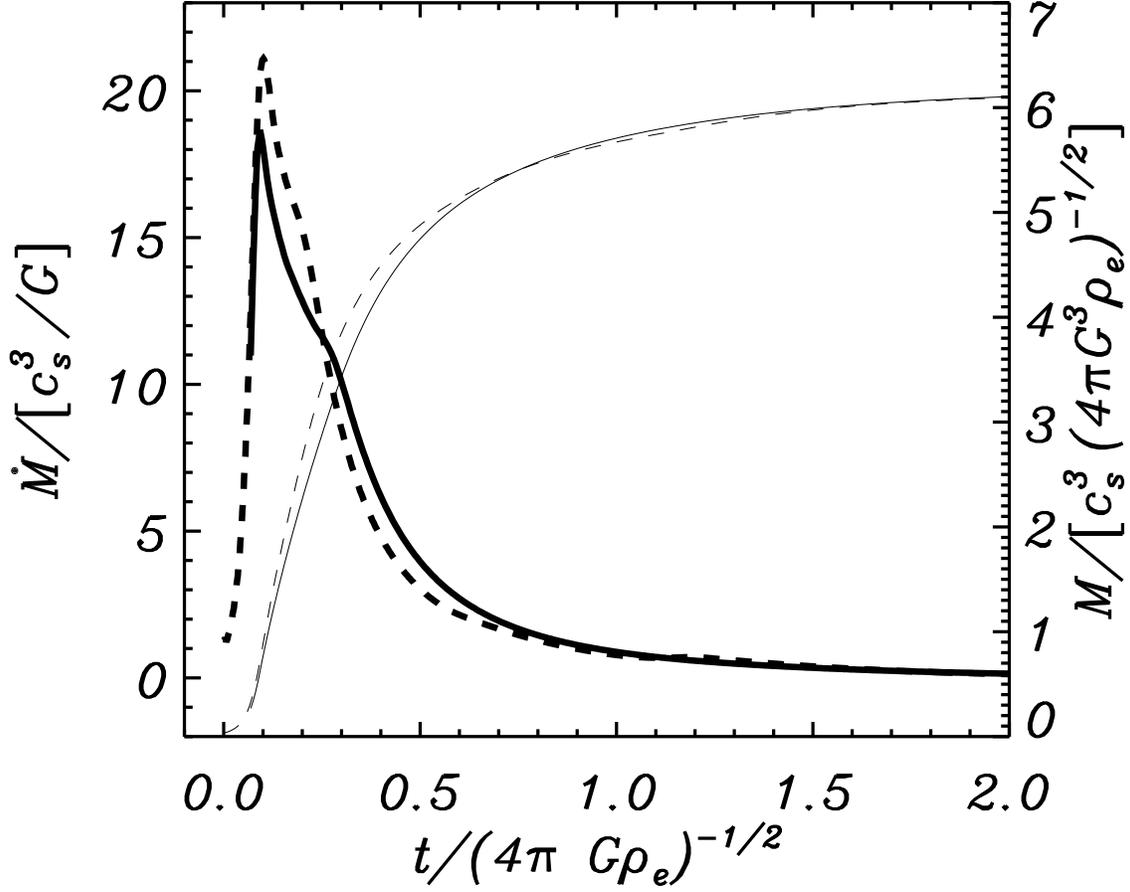}
}
\caption{Comparison of accretion for 3D Cartesian simulation with 1D spherically-symmetric
simulation, for evolution of Bonnor-Ebert sphere. Thick curves, left scale: the temporal 
evolution of the mass accretion rate at the inner edge of the grid. Thin curves, 
right scale: evolution of the central point mass. The solid curves are for the 3D simulation 
with \emph{Athena} and the dashed curve is for the 1D simulation with \emph{ZEUS}.}
\label{fig:be_mass_flux}
\end{figure}

Figure \ref{fig:be_mass_flux} shows the accretion rate and the mass of the
sink particle from our 3D simulation, in comparison with the 1D simulation of \citet{gong09}
obtained with the \emph{ZEUS} code. The solid lines are from the 3D simulation and the 
dashed lines are from the 1D simulation. The accretion rates immediately after the creation of 
the particle differ, but the accretion rates at later stages are almost exactly the same. The 
peak accretion rate is higher for 1D than 3D because it is measured closer to the sink 
particle (0.5 zones vs. 1.5 zones). In addition, the point of singularity formation
cannot be resolved as well on a 3D Cartesian grid as 1D spherically-symmetric grid.
However, it is evident that the 3D code captures the overall accretion history very well.

\begin{figure}[ht]
\centerline{
\includegraphics[width=0.9\textwidth]{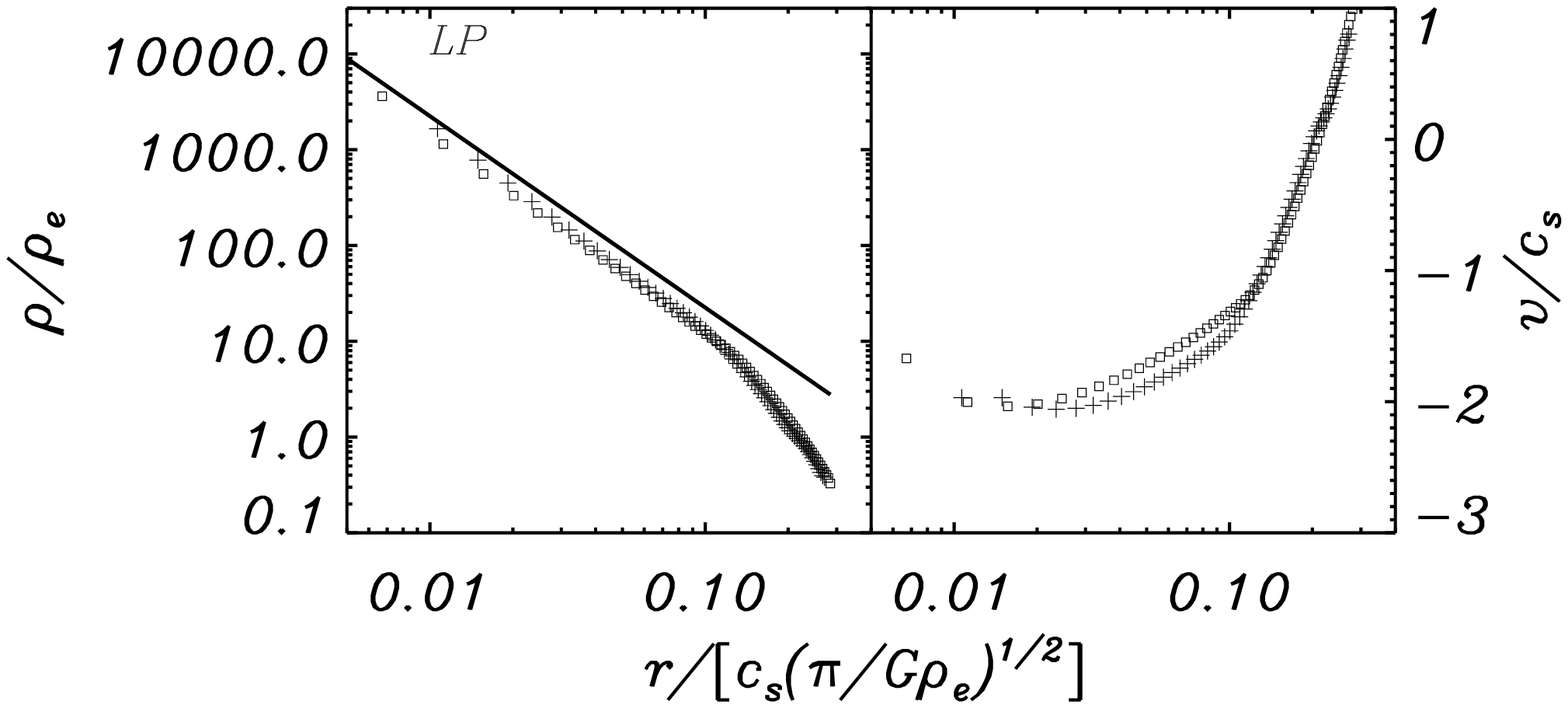}
}
\caption{Density (left) and radial velocity (right) profile comparison between 3D \emph{Athena}
simulation
and 1D spherically-symmetric \emph{ZEUS} simulation for the instant of singularity in Larson-Penston
collapse. The pluses are for the 1D simulation and the squares are for the 3D simulation.
Also shown is the singular LP density profile (Equation (\ref{lp_den}); solid line on left).
Units are normalized using the initial density at the edge of the sphere, $\rho_e$. The 
initial outer radius of the sphere is $1.1 r_{\rm BE,crit}$, and the initial density in the
region exterior to the sphere is $0.001 \rho_e$.}
\label{fig:be_rhov_comp}
\end{figure}

To further check the results of the model evolution, Figure \ref{fig:be_rhov_comp} compares
the density and velocity profiles for 3D and 1D simulations at the instant of the sink 
particle creation. The density profiles both approach the Larson-Penston singular solution
(Equation (\ref{lp_den})), while the velocity profiles approach the limit $v = -3.4 c_s$, 
in the inner part.

\begin{figure}[ht]
\centerline{
\includegraphics[width=0.9\textwidth]{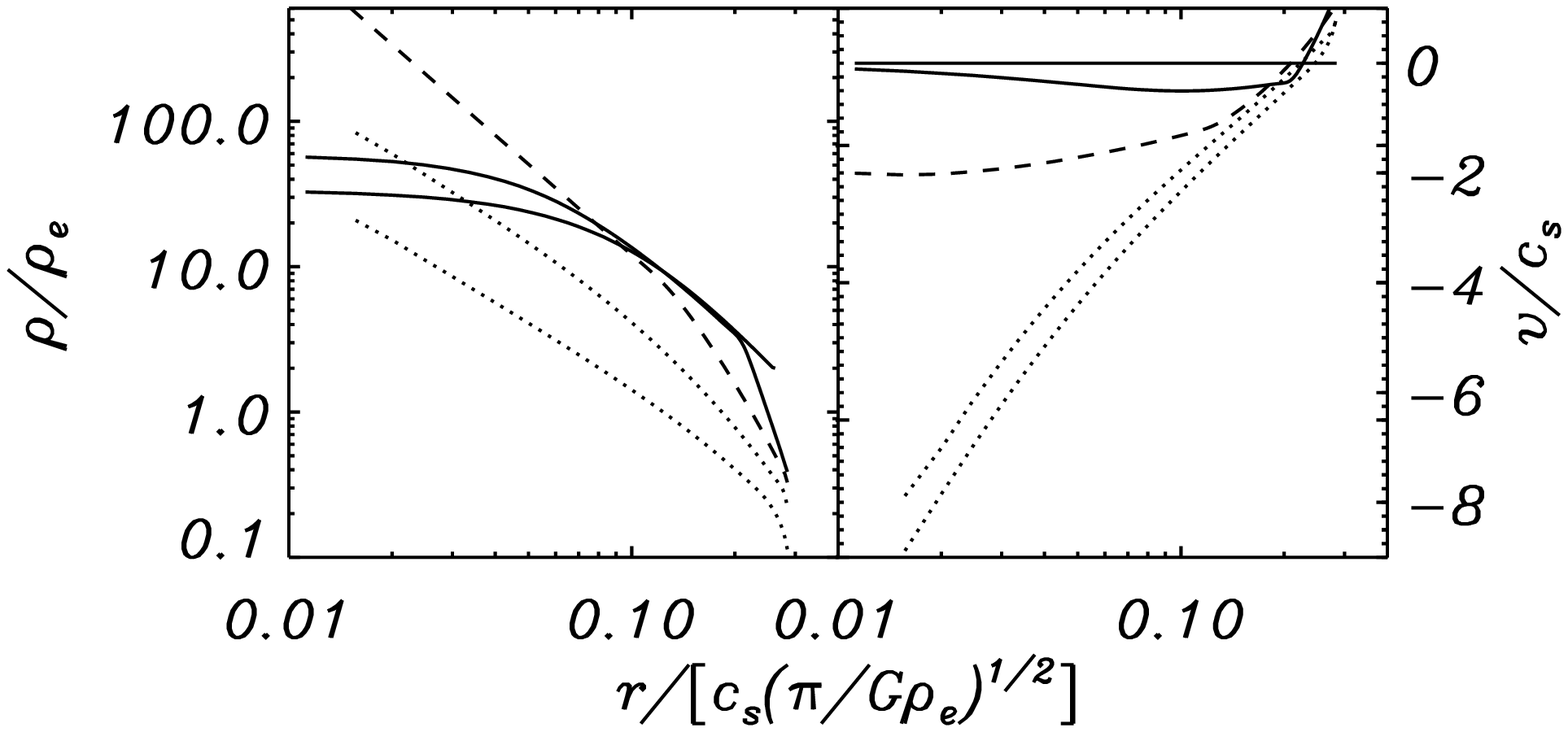}
}
\caption{Radial density (left) and velocity (right) profiles throughout collapse and infall stages for
test beginning with static Bonnor-Ebert density profile. 
The solid lines are during collapse 
($t=0$ for the lower and $t= 0.043 t_0$ for the upper curve in the left panel, and the opposite in the 
right panel), the dashed lines are at the instant of sink particle creation, 
and the dotted lines are during the infall stage ($t=0.129 t_0$ for the upper and $t=0.172 t_0$ 
for the lower curve in each panel). The time unit is $t_0= c_s(\pi/G\rho_e)^{1/2}$
for $\rho_e$ the initial density at the edge of the sphere. The behavior in each stage
is consistent with previous results from spherically-symmetric simulations \citep{gong09}.
The time interval between profiles is $\Delta t = 0.043 t_0$.}
\label{fig:be_3d_rhov_profile}
\end{figure}

Figure \ref{fig:be_3d_rhov_profile} shows the radial density and velocity evolution for
the 3D simulation for this test. The solid lines show the profiles before the sink particle 
is created, the dashed lines show the profiles at the instant of the sink particle creation,
and the dotted lines show the profiles during the envelope infall stage \citep[see][]{gong09}.
Profiles are separated by time $\Delta t = 0.043 (\pi/G \rho_e)^{1/2}$. The collapse of the 
inner part (and the expansion of the outer part) is clearly seen.
The density approaches the singular LP profile $\rho \propto r^{-2}$ (Equation (\ref{lp_den}))
at the instant of collapse, and the corresponding velocity profile is flat in the inner part, 
approaching $-3.4 c_s$. After the creation of the sink particle, the inner density and velocity 
profiles approach free-fall.

\subsection{Converging Supersonic Flows}
\citet{gong09} presented a unified model for dense molecular core formation and evolution,
based on spherically-symmetric simulations. In that work, dense cores are not present
as either stable or unstable density concentration in the initial conditions, but are built 
by the convergence of supersonic flows. Post-shock compressed gas accumulates
over time in stagnant, shock-bounded regions. When the core accumulates enough mass, it
becomes gravitationally supercritical and collapses, leading to formation of a protostar
at the center. This is followed by a stage of infall of the envelope onto the protostar,
and subsequent accretion of ambient material.

To ensure the accretion rate of sink particles is accurate for cores formed by supersonic
flows, we run a simulation of a 3D spherical converging supersonic flow, and compare to the 
accretion rate obtained by \citet{gong09} using a 1D spherically-symmetric simulation.
The initial density is uniform everywhere, with a value $\rho_0$ (this represents a typical
density within GMCs). The flow in the initial conditions converges to the center everywhere 
at Mach $2$. The size of the simulation box is $1.6 c_s/(4 \pi G \rho_0)^{1/2}$. This size
is chosen so that it is large enough for the post-shock compressed region to grow until
it collapses. The simulation is run with $129^3$ cells.
For the 1D spherical symmetric simulation, the initial density and velocity profiles
are the same as 3D but the model is run with $64$ zones. For the 1D simulation, a sink cell
is introduced at the center after collapse occurs, with an outflow boundary condition 
next to the sink cell \citep{gong09}.

\begin{figure}[ht]
\centerline{
\includegraphics[width=0.9\textwidth]{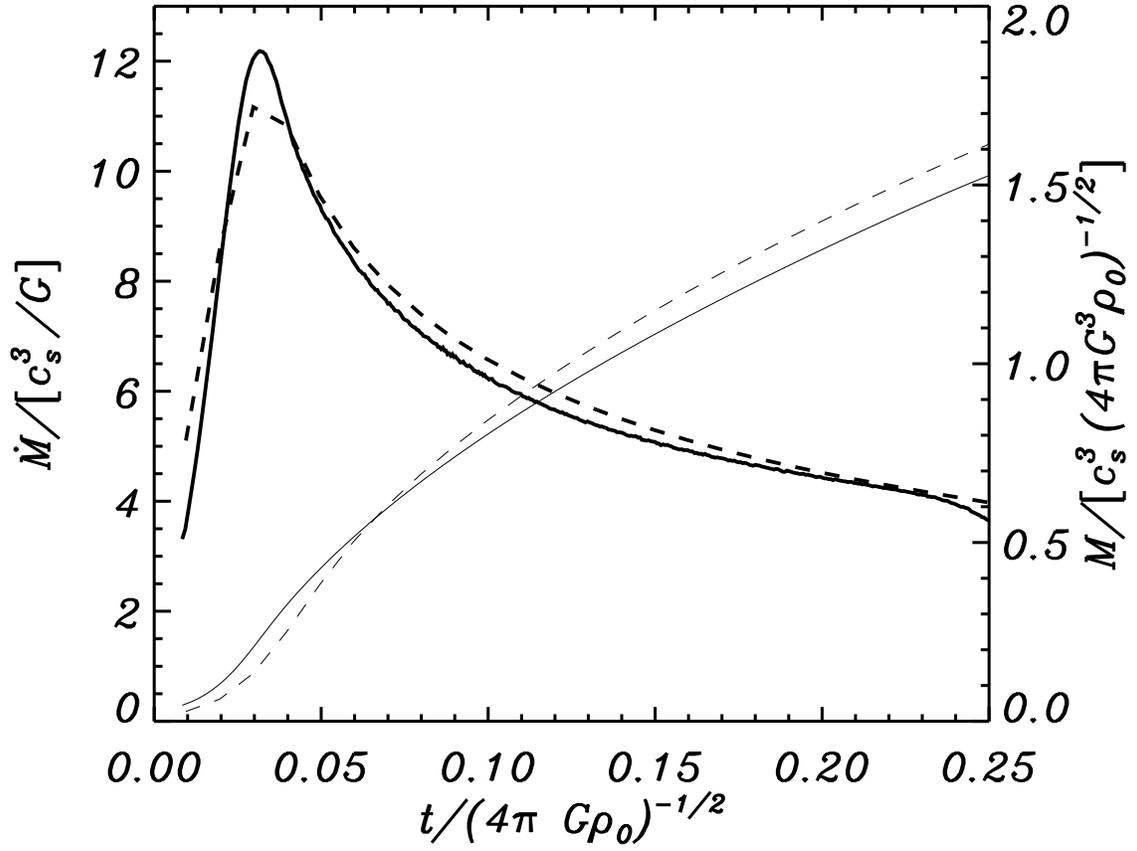}
}
\caption{Comparison of accretion evolution for 3D Cartesian and for 1D spherically-symmetric 
simulation for spherical converging supersonic (Mach 2) flow. Heavy curves show accretion rate,
and light curves show sink particle mass. Solid curves are the 3D model with \emph{Athena}, and 
dashed curves are the 1D model with \emph{ZEUS}.}
\label{fig:cvg_mass_flux}
\end{figure}

Figure \ref{fig:cvg_mass_flux} shows the accretion rate and the mass of the protostar 
versus time during the envelope infall stage, comparing the 3D and 1D simulation results.
The accretion rate is nearly same for the two cases, and the mass history of the ``protostar'' 
is comparable. For the 1D simulation, the accretion rate will decrease to exactly the inflow
value $8\pi G\rho_0 r_{\rm box}$ at late time. For 3D simulation, however, we cannot create 
an ideal spherical converging flow given our cubic domain, such that the accretion rate 
decreases after $t \sim 0.2 (4 \pi G \rho_0)^{-1/2}$. Note that the peak of the accretion 
rate for the 1D model is smaller than the value in \citet{gong09} for Mach $2$, due to
lower resolution here.

\section{Planar Converging Supersonic Flow with Sink Particles}
The most generic configuration for converging supersonic flow is planar, and 
\citet{gong11} presented a set of 3D numerical simulations with this geometry -- combined with
turbulent perturbations -- to study the core building and collapse.
The models of \citet{gong11} represent a localized region within a giant molecular cloud, 
in which there is an overall convergence in the velocity field (produced by the largest-scale
motions in the cloud), combined with a turbulent power spectrum in which linewidth increases
with size. Since sink particle techniques were not employed in \citet{gong11}, each model
simulation was stopped when the most evolved core collapsed. To demonstrate the capabilities
for following multiple core collapse and evolution when sink particles are introduced, here we 
rerun a sample simulation from the \citet{gong11} suite.

As in \citet{gong11}, we adopt the isothermal approximation. The isothermal sound speed at a 
temperature
$T$ is
\begin{equation}\label{isosp_def}
c_s = 0.20 \kms \left(\frac{T}{10\K}\right)^{1/2}.
\end{equation}
If the density within clouds were uniform, the spatial scale relevant
for gravitational instability would be the Jeans length
\begin{equation}\label{LJ_def}
L_J \equiv c_s \left(\frac{\pi}{G \rho_{0}}\right)^{1/2} =2.76 \pc
\left(\frac{n_{\rm H,0}}{10^2\pcc}\right)^{-1/2}
\left(\frac{T}{10\K}\right)^{1/2},
\end{equation}
evaluated at the mean density $\rho_0$. The corresponding Jeans mass is
\begin{equation}\label{MJ_def}
M_J \equiv \rho_0 L_J^3 = c_s^3 \left(\frac{\pi^3}{G^3\rho_0}\right)^{1/2}
=72
\Msun \left(\frac{n_{\rm H,0}}{10^2 \pcc }   \right)^{-1/2}  
\left(\frac{T}{10\K  }  \right)^{3/2}.
\end{equation}
The Jeans time at the mean cloud density is
\begin{equation}\label{tJ_def}
t_J \equiv \frac{L_J}{c_s} = \left(\frac{\pi}{G \rho_0}\right)^{1/2}=3.27\, t_{\mathrm{ff}}(\rho_0)=
1.4\times 10^7 \yr\, \left(\frac{n_{\rm H,0}}{10^2\pcc}\right)^{-1/2}.
\end{equation}
Since we consider a planar converging flow, another relevant quantity is 
the total surface density integrated along the $z$ direction,
\begin{equation}\label{surfd_def}
\Sigma = \int \rho(x,y,z) dz = \Sigma_0 \int \frac{\rho}{\rho_0} \frac{dz}{L_J},
\end{equation}
for $\Sigma_0 \equiv \rho_0 L_J = 9.49 \Msun \pc^{-2} 
(T/10K)^{1/2}(n_{H,0}/10^2 \pcc)^{1/2}$.

For this test, the supersonic flow converges to the central plane from $+z$ and $-z$ directions 
at Mach number $\mathcal{M} = 5$, for total relative Mach number 10. For both the whole domain 
initially and the inflowing gas subsequently, we apply perturbations following a Gaussian 
random distribution, with a Fourier power spectrum of the form
\begin{equation}\label{pert_pow_eq}
  \langle \left| {\delta \textbf{\em v}_k} \right|^2 \rangle \propto k^{-2},
\end{equation}
for $|kL/2 \pi| < N/2$, where $N$ is the resolution and $L$ is the
size of the simulation box in $x$ and $y$. The power spectrum is
appropriate for supersonic turbulence as observed in GMCs
\citep{mcke07}. The perturbation velocity fields are pre-generated
with resolution $256^3$ in a box of size $L_J^3$. The perturbation
fields are advected inward from the $z$-boundaries at inflow speed
$\mathcal{M}\,c_s$: at time intervals $\Delta t = \Delta
z/(\mathcal{M}c_s)$, slices of the pre-generated perturbation fields
for $v_x, v_y$ and $v_z$ are read in to update values in the ghost
zones at the $z$-boundaries. 

We set the amplitude of the turbulent power spectrum at the large scale to 
$\delta v_{1D}(L_J)=1.3 c_s$, which corresponds to the low amplitude case in \citet{gong11}.
The resolution is $N_x \times N_y \times N_z = 256 \times 256 \times 96$,
with domain size $L_x \times L_y \times L_z /L_J^3= 1 \times 1 \times 0.375$.

\subsection{Structure Evolution}
\begin{figure}[ht]
\centerline{
\includegraphics[width=0.9\textwidth]{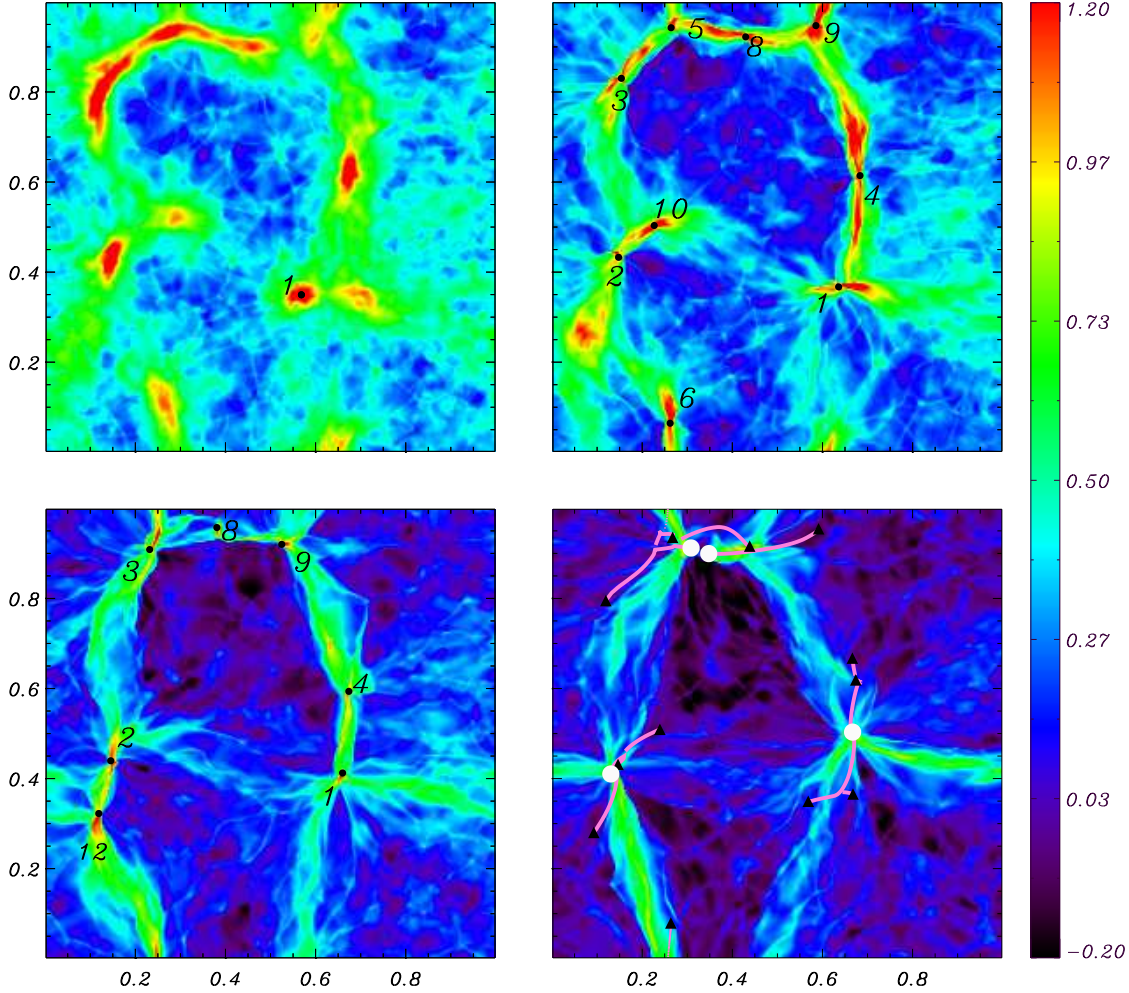}
}
\caption{Evolution of surface density projected in the $z$-direction (color scale $
{\rm log}\Sigma/\Sigma_0$) for a planar converging flow simulation with inflow Mach 
number $\mathcal{M} =5$, and supersonic turbulence. The four panels from top left to bottom 
right show snapshots at four instants: $0.301 t_J$, $0.349 t_J$, $0.398 t_J$, and $0.446 t_J$. 
The top left panel shows the surface density when the first sink ``1'' is created. The 
numbers in panels mark the time sequence of sink formation. The black solid dots in
the first three frames show instantaneous locations of sink particles. In the final panel,
the four white large solid dots show the surviving sinks after mergers. The
magenta curves shows the trajectories of all sinks from birth (marked with triangles) 
to the end of the simulation. The area of the projected domain is $L_J \times L_J$ (see
Equation (\ref{LJ_def})).}
\label{fig:protostar_formation}
\end{figure}

Figure \ref{fig:protostar_formation} shows evolution of the surface density 
(Equation (\ref{surfd_def})) projected in the $z$-direction after the most
evolved core collapses and creates a sink particle (marked as ``1'' in the images). 
The instants for the four images from top left to the bottom rights are: 
$0.301 t_J$, $0.349 t_J$, $0.398 t_J$ and $0.446 t_J$. The time interval between 
these images is 
\begin{equation}\label{dt}
\Delta t = 0.048 t_J = 6.72 \times 10^5 \yr \left(\frac{n_{\rm H,0}}{10^2 \pcc}\right)^{-1/2}. 
\end{equation}
The black dots and numbers mark the sink particles formed prior to the time of each
snapshot for the first three plots. In the last frame, the magenta curves show the trajectories
of sink particles and the black triangles show where these sinks were created. Over time,
some of the sinks merge, and the large white solid dots show the final set of post-merger 
sink locations at $t=0.446t_0$. The numbers marking sinks in the images indicates the sequence 
of their creation. In the top right figure, the sink number 7 is missing since it has merged 
with the sink 1. Similarly, in the lower left, several of the sinks have already undergone
mergers, and their corresponding numbers are not shown (5, 6, 7, 10, 11).

Filamentary features dominate the moderate--density structure in all images. These structures
grow from the initial turbulent perturbations, which are then amplified by self-gravity 
(see Fig.1 in \citet{gong11}). The localized collapse of these filamentary structures 
leads to the formation of protostars. Filaments also become more stratified over time as
they acquire more material and contract perpendicularly under their self-gravity.

The bottom right panel in Figure \ref{fig:protostar_formation} shows the trajectories of all 
sinks, as well as their merging history. There were 12 sinks created during this simulation,
and four survive at final time. All of the other 8 have merged with other nearby sinks
due to close encounters.

\subsection{Sink Particle Mass Evolution}
\begin{figure}[ht]
\centerline{
\includegraphics[width=0.9\textwidth]{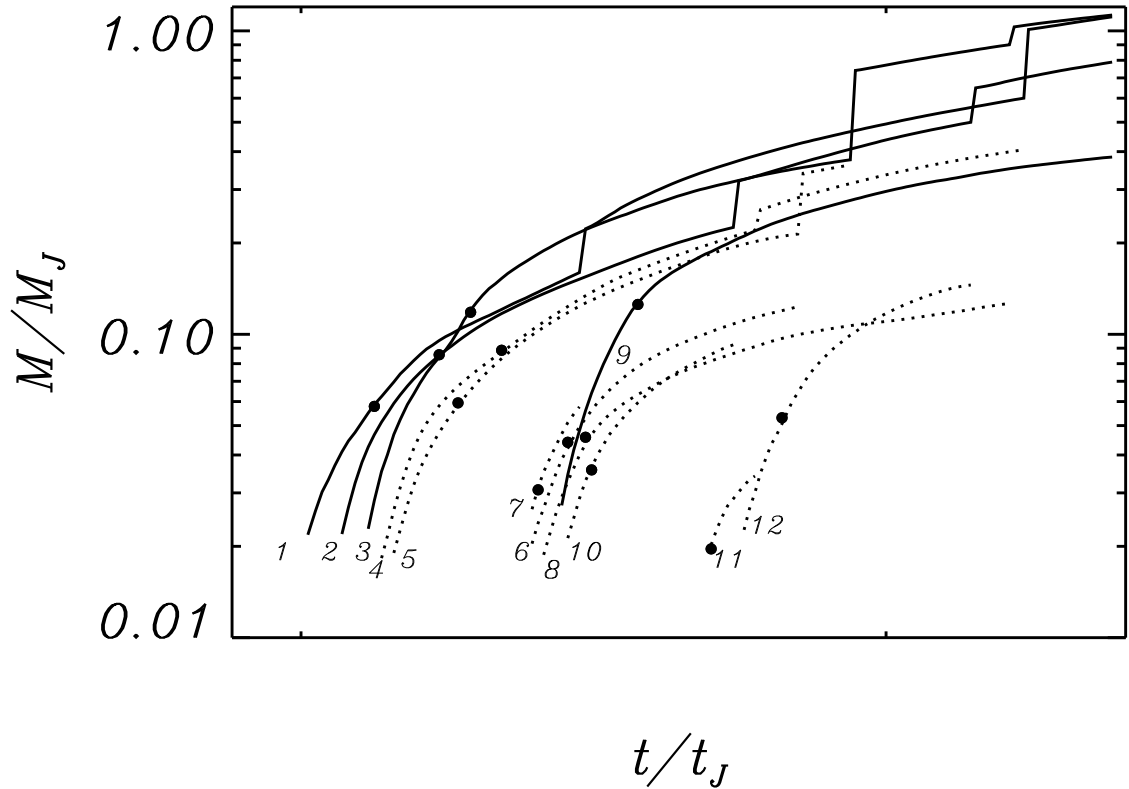}
}
\caption{Temporal evolution of sink particle masses (see Equation (\ref{MJ_def}) and
Equation (\ref{tJ_def}) for definition of $M_J$ and $t_J$ respectively). The solid black lines
show the sinks that survive to late times. The dotted lines show the sinks
that are eventually merged into larger particles. The solid black dots show the
gravitationally bound dense core masses immediately before the creation of the
corresponding sink particle at its center.}
\label{fig:mass_t}
\end{figure}

Figure \ref{fig:mass_t} shows the mass of sinks versus time. The solid lines
show the four sinks that survive up until late stages. The dotted lines are sinks
that undergo mergers with other more massive sinks. Each sink forms at the center of a dense
core, and the solid dots show the gravitationally bound core masses calculated using the 
GRID-core finding algorithm \citep{gong11} immediately before the formation 
of the corresponding sink particle. The mass of every sink particle grows smoothly to 
reach and exceed the bound core mass, and keeps increasing until it merges with other sinks.
As indicated in the figure, the final masses of sink particles are all much larger than
the bound core masses at the initial instant of sink formation, which are in the range 
$m_{\rm core} = 0.02 - 0.1 M_J$.

The formation of sinks is divided into three groups. Protostars 1, 2, 3, 4 and
5 form at the earliest time; sinks 6, 7, 8, 9 and 10 form a bit later; and sinks 
11 and 12 form during the late accretion stage. Sinks forming at earlier stages
are more likely to survive to the end, and their masses
 grow significantly via mergers and
late accretion.
For example, the merging history for protostar 1 is: $7\rightarrow1\leftarrow4\leftarrow11$,
the merging history for protostar 2 is: $10\rightarrow 2 \leftarrow 12$, and the merging
history for protostar 3 is: $6 \rightarrow 5 \rightarrow 3 \leftarrow 8$.

At late stages, the masses of these surviving sink particles are very high: $M_1 = 1.11 M_J$,
$M_2 = 0.79 M_J$, $M_3 = 1.13 M_J$ and $M_9 = 0.38 M_J$. These correspond to tens of $\Msun$ 
if $n_{\rm H,0} =10^2 - 10^3 \pcc$. In part, these sinks may end up with very high mass 
because the turbulence has low amplitude and is purely decaying. As consequence, matter is 
not prevented from accreting into the potential
wells that develop. In reality, outflows, radiative feedback, and other energy injection would
limit gas accretion onto protostars. In addition, some stars could be ejected due to close
gravitational encounters, before acquiring a high mass. These physical issues will be
addressed in a future publication; it is straightforward to implement localized feedback with
rates set by the mass, age, accretion rate, etc, of each sink. Here, our goal is simply to 
test the proper implementation of sink particle algorithms and to demonstrate that these 
enable robust long-term evolution.

We note that for the present algorithm (and similarly for other sink particle implementations),
the minimum particle mass depends on the grid resolution, on the minimum density threshold for
sink creation, and on the size of the sink particle control volume. The sink particle density
threshold we adopt here is $\rho_{th} = \rho_{\rm _{LP}} (0.5 \Delta x) = 8.86 c_s^2/(\pi G
(\Delta x)^2)$. With resolution 
$\Delta x = L_J/N$, $\rho_{\rm th} = 8.86 \rho_0 (N/\pi)^2$,
and the mass within the central cell of the
sink region is $0.89 M_J/N = N^{-1} 65 \Msun (n_{\rm H,0}/10^2)^{-1/2} (T/10 K)^{3/2}$. The
total initial sink particle mass are larger, due to non-negligible density in the surrounding 
sink cells. We find (see Figure \ref{fig:mass_t}) that the initial sink masses in this test 
simulation are in the range $0.02 M_J - 0.1 M_J$. With $N=256$, this ranges from 
$\sim 6 - 30$ times the minimum mass.

\subsection{Criteria for Sink Particle Creation}
\begin{figure}[ht]
\centerline{
\includegraphics[width=0.9\textwidth]{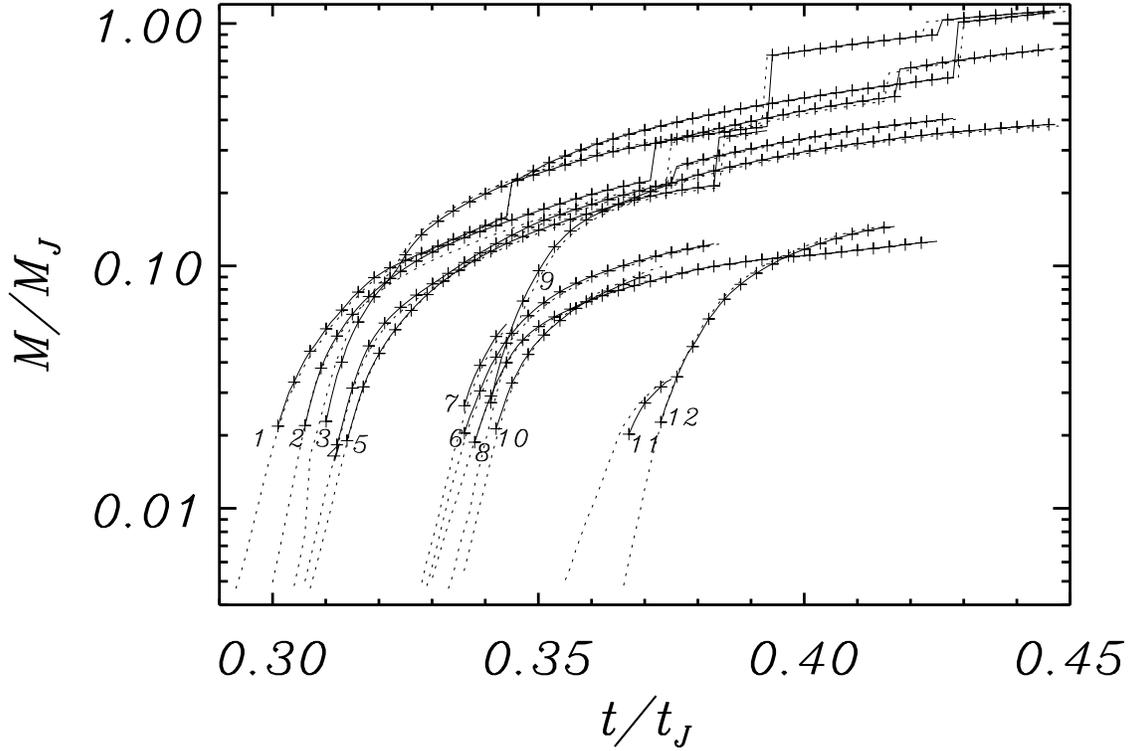}
}
\caption{Temporal evolution of sink particle masses based on adoption of different creation 
criteria. The solid lines are for sink particles created with the LP density threshold,
and local potential minimum criteria. The tracks marked by pluses are for sink particles 
created based on additional criteria: a converging flow and a gravitationally bound state. 
The dotted lines are for sink particles created based on the Truelove density threshold 
and the local gravitational potential minimum check.}
\label{fig:mass_comp}
\end{figure}

As discussed in Section 2.2.1, different density thresholds have been adopted for sink particle
creation by different groups, motivated by different physical and computational considerations.
Figure \ref{fig:mass_comp} shows a comparison of sink particle mass versus time for different 
sink particle creation criteria. The solid lines are based on criteria of exceeding 
the LP density threshold (Equation (\ref{lp_rho})) and satisfying the local potential 
minimum check. The tracks marked by pluses adopt additional criteria: the converging flow check
and gravitationally bound state check. The dotted lines adopt the Truelove density instead of
the (higher) LP density as a threshold, and apply a check for a local gravitational potential
minimum. We find no differences between the sink masses for the first two sets of criteria.
For the third set of criteria, the sink masses at birth are much lower than in the first case,
because of the lower density threshold. Compared to initial sink masses of $\sim 0.02 M_J$ using
the LP density criterion, initial sink masses are $\sim 0.005 M_J$ using the Truelove density
criterion. However, these sinks evolve to follow tracks identical to those found with the other
two sink criteria choices. Analogous comparisons at different Mach numbers show that no 
additional particles are created by artificial fragmentation when using our standard criteria, 
even though the LP density threshold is higher than the Truelove density threshold.

We conclude that a proper density threshold and local potential minimum check are probably
sufficient criteria for sink particle creation. By choosing a density threshold given by the LP
solution, and requiring that the zone is at a local potential minimum, we can ensure that the 
region surrounding a cell is under gravitational collapse.

We note that in the study of \citet{true97}, there was no implementation
of sink particles. There, the authors showed that artificial fragmentation
may occur in AMR simulations if the ratio of the Jeans length to the grid scale
is too small. However, in simulations where sink particles are introduced,
the formal requirements on resolution need not be identical to the case when
there are no sink particles. Previous authors \citep{krum04,fede10} showed that they
were able to avoid artificial fragmentation by capping the density at the
Truelove value (see Equation \ref{rho_tr}). However, we find that when a slightly higher 
density criterion  (motivated by LP collapse) is adopted for sink particle creation, 
there is still no artificial fragmentation, and in fact the mass evolution follows similar 
tracks as it does when the Truelove density threshold is used (see Fig. \ref{fig:mass_comp}). 
The similarity among the mass
evolution tracks for 
different sink particle creation criteria suggests that the class of 
sink particle methods currently in use provides reliable results for 
core collapse and accretion.

\section{Summary}
We have presented an implementation of sink particles to the grid-based Eulerian hydrodynamics
code \emph{Athena}. A standard particle-mesh method is adopted to calculate gravity forces by
and on the sink particles, with the Poisson equation solved via FFT methods. 
We use the mass and momentum fluxes from the Riemann solver to update the mass and momentum of 
sink particles. Criteria for sink particle creation are similar to those used by other authors,
although we suggest that a higher density threshold is motivated by the Larson-Penston profile
that is known to develop as a generic stage of self-gravitating collapse. The Larson-Penston density
threshold we adopt for sink creation in our method is a factor 14 larger than the ``Truelove'' 
density threshold adopted in other sink implementations; our method also differs from other
implementations in that the sink region surrounding each particle consists of ghost zones
rather than active zones. Outside the sink region, the density is below the Truelove value.
Our tests show similar results whether we use higher or lower density thresholds for sink
formation; additional testing in the future can explore whether artificial fragmentation
is avoided in all cases. We validate our 
method and implementation with a series of tests. These tests include comparison of accretion
rates with analytic solutions for self-similar collapse of isothermal spheres, and comparison
of accretion rates and solution profiles with 1D spherically-symmetric collapse of Bonnor-Ebert
spheres and cores formed by converging supersonic flows. We demonstrate Galilean invariance
of our accretion solutions onto sink particles.

To demonstrate application of our method, we present sample results for a simulation
of planar converging supersonic flows with turbulence. Filaments form, and local collapse
produces sink particles which then accrete material from filaments. The mass smoothly increases
to exceed the mass of the initial bound core in which a sink formed. Sinks forming at early 
stages merge with other smaller sinks if sink regions (of standard size $3^3$ zones) overlap.
In the context of these converging flow tests, we have investigated various criteria 
that have been adopted 
for creation of sink particles. We find that a (large enough) density 
threshold and local gravitational potential minimum check give the same results as more strict
sets of criteria in this case (and in other tests we have conducted).

The implementation of sink particles we present enables study of a wide range of astrophysical
applications involving gravitational collapse of a gaseous medium. In particular, robust and
accurate methods for implementing sink particles make it possible to track 
sequential formation of multiple protostars over long periods under realistic environmental
conditions. As such, sink particles represent a crucial numerical tool for addressing key
unsolved problems in star formation, including the origin of the stellar initial mass function.
The implementation of sink particles we describe here will be made available to the community 
in an upcoming release of the \emph{Athena} code.

\acknowledgements
We are grateful to Aaron Skinner for his contributions to code development for particle
implementation, to Christoph Federrath for helpful discussions, 
and to the referee for 
comments that helped us to improve the manuscript.
This work was supported by grant NNXI0AF60G from NASA's Astrophysics Theory Program. 
Computations made use of the HPCC deepthought cluster administrated by the OIT at the 
University of Maryland, as well as the Borg cluster in the Department of the Astronomy.

\appendix

\section*{APPENDIX: VACUUM BOUNDARY CONDITION POTENTIAL VIA FOURIER TRANSFORMS}
In this section, we provide details for the solution of Poisson's equation with vacuum boundary
conditions. Let $-4 \pi G \mathcal{G}(\textbf{x},\textbf{x}')$ be the Green's function solution for the 
Poisson equation $\nabla^2 \Phi = -4 \pi G \delta (\textbf{x}-\textbf{x}')$, so that the 
potential produced by a density field $\rho(\textbf{x})$ is:
\begin{equation}\label{phi_x}
\Phi(\textbf{x}) = -4 \pi G \int \mathcal{G}(\textbf{x},\textbf{x}') \rho(\textbf{x}') d^3\textbf{x}.
\end{equation}
For potential $\Phi$ at ($x_a, y_b, z_c$) within a domain $(L_x, L_y, L_z)$ with dimensions 
$(N_x, N_y, N_z)$, the corresponding discrete sum is:
\begin{equation}\label{phi_xyz}
\Phi(x_a,y_b,z_c) = -4 \pi G \sum_{l=0}^{N_x-1} \sum_{m=0}^{N_y-1} \sum_{n=0}^{N_z-1} 
   \mathcal{G}(x_a-x_l,y_b-y_m,z_c-z_n) \rho(x_l,y_m,z_n);
\end{equation}
here $(a,b,c)$ and $(l,m,n)$ are the integer indices for the corresponding coordinates $x,y,z$
respectively. $\Phi(x_a,y_b,z_c)$ is a convolution of 
$\mathcal{G}(\textbf{x})$ and $\rho(\textbf{x})$.  

Because $\mathcal{G}(|\textbf{x}-\textbf{x}'|)$ is a function in the domain $[-L_x,L_x]\times[-L_y,L_y] \times[-L_z,L_z]$, and $\rho(\textbf{x}')$ is a 
function in the domain $[0,L_x]\times[0,L_y]\times[0,L_z]$, if we define
$\rho(x_l,y_m,z_n) = 0$ for $l<0, m<0$, or $n <0$, Equation (\ref{phi_xyz}) can be re-written 
as:
\begin{equation}\label{phi_xyz_n}
\Phi(x_a,y_b,z_c) = -4 \pi G \sum_{l=-N_x}^{N_x-1} \sum_{m=-N_y}^{N_y-1} \sum_{n=-N_z}^{N_z-1} 
   \mathcal{G}(x_a-x_l,y_b-y_m,z_c-z_n) \rho(x_l,y_m,z_n),
\end{equation}

We may define periodic functions with period $2L_x, 2L_y$ and $2L_z$ in $x, y$ and $z$
directions respectively, such that $\rho(\textbf{x}')$ and $\mathcal{G}(\textbf{x},\textbf{x}')$ agree
with these periodic functions for $x \in [-L_x,L_x], y \in [-L_y, L_y]$ and
$z \in [-L_z, L_z]$. Then from the Fourier convolution theorem, Equation (\ref{phi_xyz_n}) 
can be expressed in terms of the respective transforms $\hat{\mathcal{G}}$ and $\hat{\rho}$
of $\mathcal{G}(\textbf{x},\textbf{x}')$ and $\rho(\textbf{x}')$ from the Fourier
convolution theorem:
\begin{equation}\label{phi_conv}
\Phi(x_a,y_b,z_c) = \frac{-4 \pi G}{(2N_x)(2N_y)(2N_z)}
   \sum_{i=0}^{2N_x-1} \sum_{j=0}^{2N_y-1} \sum_{k=0}^{2N_z-1} 
   \hat{\mathcal{G}}_{ijk}  \hat{\rho}_{ijk} 
   e^{-2\pi i\left(\frac{ai}{2N_x}+\frac{bj}{2N_y}+\frac{ck}{2N_z}\right)},
\end{equation}
Note the summation index for $i$ can be either from $-N_x \to N_x$ or $0 \to 2N_x-1$ because 
$\hat{\mathcal{G}}$ and $\hat{\rho}$ are periodic. The same applies to $j$ and $k$ indices.
In our implementation within \emph{Athena}, Equation (\ref{phi_conv}) is further decomposed 
into a sum over even and odd terms to save memory.
{}

\newpage

\end{document}